\documentclass[
 aip,
 apl, 
 amsmath,amssymb,
 reprint,%
floatfix
]{revtex4-2}
\preprint{AIP/123-QED}
\usepackage{bm}              
\usepackage[utf8]{inputenc}
\usepackage[T1]{fontenc}
\usepackage{mathptmx}
\usepackage{graphicx}
\usepackage{float}
\usepackage[separate-uncertainty=true]{siunitx}        
\usepackage[colorlinks]{hyperref}       
\usepackage{import}
\usepackage[capitalise,nameinlink]{cleveref}
\newcommand{\crefadd}[2]{\hyperref[#1]{\cref{#1}#2}} 
\newcommand{\Crefadd}[2]{\hyperref[#1]{\Cref{#1}#2}} 
\usepackage{xcolor}               
\definecolor{navygray}{RGB}{140,150,200}
\hypersetup{
citecolor={navygray},
linkcolor={navygray},
urlcolor={navygray},
}

\begin{document}
\title{High quality superconducting tantalum resonators with beta phase defects}

\author{Ritika~Dhundhwal}
\affiliation{IQMT, Karlsruhe Institute of Technology, 76131 Karlsruhe, Germany}

\author{Haoran~Duan}
\affiliation{INT, Karlsruhe Institute of Technology, 76131 Karlsruhe, Germany}

\author{Lucas~Brauch}
\affiliation{IQMT, Karlsruhe Institute of Technology, 76131 Karlsruhe, Germany}
\author{Soroush~Arabi}
\affiliation{IQMT, Karlsruhe Institute of Technology, 76131 Karlsruhe, Germany}
\affiliation{PHI, Karlsruhe Institute of Technology, 76131 Karlsruhe, Germany}
\author{Dirk~Fuchs}
\author{Amir-Abbas~Haghighirad}
\affiliation{IQMT, Karlsruhe Institute of Technology, 76131 Karlsruhe, Germany}

\author{Alexander~Welle}
\affiliation{IFG, Karlsruhe Institute of Technology, 76131 Karlsruhe, Germany}
\affiliation{KNMFi, Karlsruhe Institute of Technology, 76131 Karlsruhe, Germany}

\author{Florentine~Scharwaechter}
\author{Sudip~Pal}
\author{Marc~Scheffler}
\affiliation{Physics Institute 1, University of Stuttgart, 70569 Stuttgart, Germany}

\author{Jos\'e~Palomo}
\author{Zaki~Leghtas}
\affiliation{Laboratoire de Physique de l’Ecole normale supérieure, 75005 Paris, France}

\author{Anil~Murani}
\affiliation{Alice \& Bob, 53 Bd du Général Martial Valin, Paris 75015, France}

\author{Horst~Hahn}
\affiliation{IQMT, Karlsruhe Institute of Technology, 76131 Karlsruhe, Germany}
\affiliation{INT, Karlsruhe Institute of Technology, 76131 Karlsruhe, Germany}

\author{Jasmin~Aghassi-Hagmann}
\affiliation{INT, Karlsruhe Institute of Technology, 76131 Karlsruhe, Germany}

\author{Christian~K\"ubel}
\affiliation{INT, Karlsruhe Institute of Technology, 76131 Karlsruhe, Germany}
\affiliation{KNMFi, Karlsruhe Institute of Technology, 76131 Karlsruhe, Germany}
\affiliation{Technische Universit\"at Darmstadt, 64289 Darmstadt, Germany}

\author{Wulf~Wulfhekel}
\affiliation{IQMT, Karlsruhe Institute of Technology, 76131 Karlsruhe, Germany}
\affiliation{PHI, Karlsruhe Institute of Technology, 76131 Karlsruhe, Germany}

\author{Ioan~M.~Pop}
\affiliation{IQMT, Karlsruhe Institute of Technology, 76131 Karlsruhe, Germany}
\affiliation{PHI, Karlsruhe Institute of Technology, 76131 Karlsruhe, Germany}
\affiliation{Physics Institute 1, University of Stuttgart, 70569 Stuttgart, Germany}

\author{Thomas~Reisinger}
\altaffiliation{thomas.reisinger@kit.edu}
\affiliation{IQMT, Karlsruhe Institute of Technology, 76131 Karlsruhe, Germany}

\begin{abstract}
For practical superconducting quantum processors, orders of magnitude improvement in coherence is required, motivating efforts to optimize hardware design and explore new materials. Among the latter, the coherence of superconducting transmon qubits has been shown to improve by forming the qubit capacitor pads from $\alpha$-tantalum, avoiding the meta-stable $\beta$-phase that forms when depositing tantalum at room temperature, and has been previously identified to be a source of microwave losses. In this work, we show lumped element resonators containing $\beta$-phase tantalum in the form of inclusions near the metal-substrate interface with internal quality factors ($Q_\text{i}$) up to \num{5.0+-2.5d6} in the single photon regime. They outperform resonators with no sign of the $\beta$-phase in x-ray diffraction and thermal quasi-particle loss. Our results indicate that small concentrations of $\beta$-phase can be beneficial, enhancing critical magnetic fields and potentially, for improving coherence in tantalum based superconducting circuits.
\end{abstract}

\maketitle

Superconducting circuits are a leading platform for quantum information processing~\cite{Blais2021cqed} and the last two decades have been marked by remarkable progress in qubit coherence times~\cite{Krasnok_2024,Krantz2019Jun}. Scaled up circuits with tens of qubits have been used to successfully demonstrate key milestones such as quantum supremacy~\cite{Quantumsupremacy2019} and error correction~\cite{sivak2023}. As a further example for the utilization of superconducting circuits, quantum-limited Josephson parametric amplifiers form key ingredients for high fidelity qubit readout and state preparation~\cite{Jose2020_JPA}. However, for solid-state macroscopic circuits, a major challenge faced by superconducting devices is decoherence~\cite{siddiqi_engineering_2021}.
In particular, dielectric loss is one of the major loss mechanisms in state-of-the-art hardware~\cite{McRae2020},
which can be mitigated both by improving circuit design or the constituent materials~\cite{deLeon2021Apr,Chayanun2024_aluminum-on-silicon,tuokkola2024methodsachievenearmillisecondenergy}.  Recently, tantalum~\cite{place2021, wang2022} has become a member of the exclusive group of materials with low dielectric loss tangent for quantum devices, challenging the performance of established platforms based on aluminum, niobium~\cite{tuokkola2024methodsachievenearmillisecondenergy,Anferov2024nb.trilayer.junction} or titanium nitride~\cite{Chang2013,Deng2023Feb_TiN}.

Record qubit lifetimes on the order of hundreds of microseconds have been demonstrated in transmon qubits composed of Ta capacitors shunting a standard Al/AlO$_x$/Al junction~\cite{place2021,wang2022}. However, the coherence in Ta-based circuits is highly dependent on deposition and fabrication process parameters. A distinctive feature of Ta thin films is the possible formation of a tetragonal $\beta$-phase~\cite{Read1965Aug}, competing with the body centered cubic (BCC) bulk $\alpha$-phase. The latter has a superconducting transition temperature ($T_\text{c}$) of $\SI{4.4}{\K}$, while the meta-stable $\beta$-phase has a significantly lower $T_\text{c}$ of $\SI{0.5}{\K}$~\cite{SCHWARTZ1972333}. It can form when depositing on substrates at near room-temperature~\cite{Navid2012Feb}. The underlying cause for the coherence improvement in Ta devices remains a matter of intensive research and is most likely multi-faceted~\cite{ganjam_surpassing_2024}. The improvement is often linked to the use of pure $\alpha$-phase Ta and then accredited to the stable, self-limiting surface oxide~\cite{McLellan2023Jul,Mustafa2024FermiLab}. However, there is a large spread in the reported loss for Ta-based devices in the literature, possibly caused by variations in film microstructure.  

\begin{figure*}
    \centering
    \includegraphics[width=2\columnwidth]{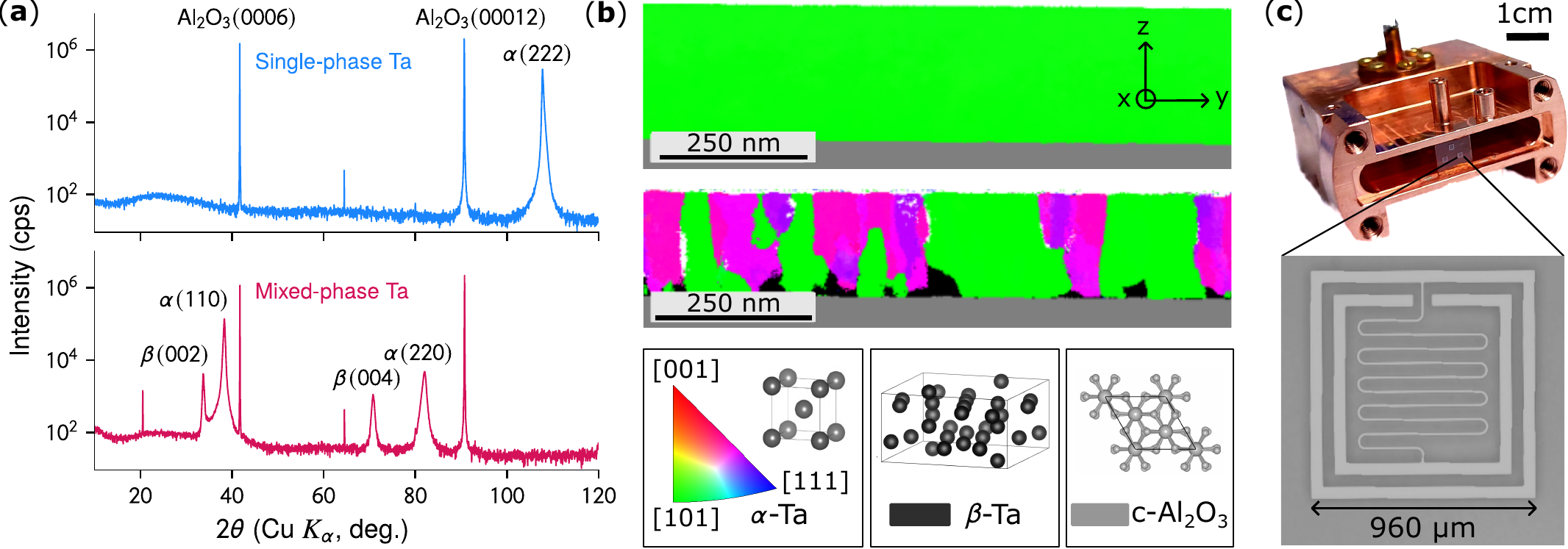}
    \caption{Crystallographic analysis of Ta films and images of resonator and 3D-waveguide used for microwave loss characterization. (a) X-ray diffractograms of Ta films grown on c-plane Al$_2$O$_3$. We use a Bruker high-resolution X-ray diffraction system in reflection, equipped with a $(022)$ Ge monochromator for the characteristic $K_\alpha$ line of a Cu X-ray source. The top panel (blue) shows a single Ta peak, denoted $\alpha(222)$, corresponding to the $\alpha$-phase growing with (111) orientation $\parallel z$ direction; the $\alpha(111)$ peak is suppressed due to the symmetry of the BCC crystal structure (Systematic absence). This film is referred to as single-phase film. The bottom panel (magenta) shows multiple Ta peaks corresponding to both $\alpha$- and $\beta$-phase growing with (110)$\parallel z$ and (001)$\parallel z$ orientation, respectively. This film is referred to as mixed-phase film. The narrow peaks visible at $\SI{20}{\degree}$ (bottom graph) and $\SI{65}{\degree}$ (both graphs) correspond to the diffraction of residual $K_\beta$ X-rays by the sapphire substrate. (b) Automated Crystal Orientation Mapping (ACOM) of lamella extracted from single (top) and mixed-phase (bottom) films performed using a transmission electron microscope. The black areas correspond to the $\beta$-phase, while the other colors (except white) indicate the crystal orientation of the $\alpha$-phase in the x-direction. The color key for the $\alpha$-phase orientation is shown in the left-most bottom panel together with a schematic of the BCC unit cell. The gray regions beneath the films indicate the sapphire substrate. Middle and right bottom panels show the unit cell for $\beta$-phase Ta and sapphire, respectively. White color areas in the mixed-phase film indicate regions where the crystal orientation is not well defined. (c) Copper 3D-waveguide sample holder (closing lid not shown) and a mounted $7\times 7$ mm$^{2}$ sapphire chip with three horseshoe style Ta resonators, similar to the one shown in the optical micrograph below it.}
    \label{fig: overview of films}
\end{figure*}

In this work, we demonstrate single photon $Q_\text{i}$ of \num{5.0+-2.5d6}
in microwave resonators fabricated from $\beta$-phase-containing Ta film. The volume fraction of the $\beta$-phase in the film is approximately $10\%$, as determined from X-ray diffraction (XRD) data - see supplemental material (SM). We show that the low power $Q_\text{i}$ of these resonators is higher than for ones composed of pure $\alpha$-phase Ta, even after aggressive post-fabrication cleaning in buffered-oxide etch (BOE)~\cite{Lozano2024Ta_on_silicon}. In order to correlate microwave loss to material properties of Ta films, we characterize the films with various material characterization tools such as XRD, four probe resistivity measurements, SQUID magnetometry, scanning tunneling microscopy (STM), scanning transmission electron microscopy (STEM) and scanning precession electron diffraction based automated crystal orientation mapping (ACOM)~\cite{Kobler2013ACOM}. We conclude that controlling the $\beta$-phase concentration, while not necessarily aiming to eliminate it, is a promising route to further reducing loss in Ta-based quantum circuits, while also enhancing critical magnetic fields ($H_{c}$). 

In~\crefadd{fig: overview of films}{(a,b)} we characterize the two types of Ta film structures, for which we aim to compare possible effects on microwave loss. The different sets of sputtering parameters and sapphire substrate cleaning procedures are summarized in the SM. First, films that exhibit only $\alpha$-phase with (111) orientation are referred to as single-phase films. The second type of film consists of both $\alpha$- and $\beta$-phase with (110) and (001) film orientation, respectively, and is referred to as mixed-phase film. We attribute the growth of the $\beta$-phase to an initially low substrate temperature and high deposition rate. The two types of films represent good examples of variation of deposition parameters leading to structurally completely different films. The orientation maps of film cross-sections shown in \crefadd{fig: overview of films}{(b)} were obtained using Automated Crystal Orientation Mapping (ACOM), a technique based on STEM~\cite{Kobler2013ACOM}, which creates a 2D array of electron diffraction patterns and correlates the patterns with a database of already simulated diffraction patterns. It generates phase and orientation maps over micron size areas. The top panel in~\crefadd{fig: overview of films}{(b)} shows an ACOM of a single-phase film lamella composed of a single crystal grain. Keeping in mind that ACOM is a highly localized probe, it still suggests that the grain size in the film is at least on the order of micrometer. The middle panel shows ACOM of a mixed-phase film lamella, confirming the presence of the $\beta$-phase: It is shown as black areas (without an indication for orientation) and is only present near the sapphire-Ta interface and not at the film surface. The $\beta$-phase may preferentially localize at the sapphire-Ta interface because of its reduced interfacial energy with the substrate. In addition, there are two main orientations of $\alpha$-phase visible leading to several grain boundaries. 

\begin{figure}
    \centering 
     \includegraphics[width=1\columnwidth]{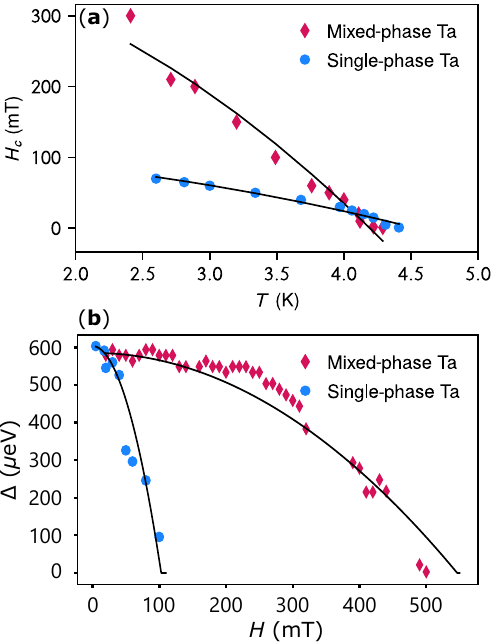}
    \caption{Critical magnetic field in parallel orientation ($H_{c}$) as a function of temperature ($T$) and superconducting gap ($\Delta$) as a function of magnetic field for the single- and mixed-phase Ta films. (a) The $H_{c}$'s were measured using a SQUID magnetometer. The black lines correspond to a fit of the data to the equation $H_c(T) = H_{c}(0)(1-(T/{T_{c}})^2)$ \cite{Bardeen1957}. (b) $\Delta$ was measured in a scanning tunneling microscope at \SI{45}{\milli\kelvin} as a function of perpendicular magnetic field after removal of the surface oxide with argon milling~\cite{Arabi2024Dec}. The black lines correspond to a fit of the data to the equation $\Delta/\Delta_{0} = 1 - (H/H_{c})^{2}$.}
    \label{fig:Tc_and_delta}
\end{figure}

The small grain size of the mixed-phase film leads to shorter superconducting coherence length and in turn a higher $H_{c}$~\cite{Hauser1964Apr}. This constitutes an additional benefit, when considering applications in magnetic fields such as hybrid spin quantum systems.~\crefadd{fig:Tc_and_delta}{(a)} presents parallel orientation measurements of $H_{c}$ of single- and mixed-phase films as a function of temperature, measured using a MPMS XL SQUID magnetometer (Quantum Design). From the fits, the zero-temperature critical fields $H_{c}(0)$ are $\SI{107+-4}{\milli\tesla}$ and $\SI{388+-17}{\milli\tesla}$ for the single- and mixed-phase films, respectively. The single-phase $H_{c}(0)$ is close to the value for bulk Ta~\cite{Hauser1964Apr}. Additionally, local measurements of the superconducting gap ($\Delta_{0}$) as a function of the magnetic field were performed with a STM~\cite{balashov_compact_2018} at $\SI{45}{\milli\kelvin}$~[\crefadd{fig:Tc_and_delta}{(b)}]. These measurements yield $\Delta_{0}$ of $\SI{600+-28}{\micro\eV}$ and $\SI{585+- 16}{\micro\eV}$ for the single- and mixed-phase films, respectively, which are only slightly smaller than the value for bulk Ta of $\SI{686}{\micro\eV}$~\cite{Reed1976Jul}. Critical magnetic fields of $\SI{103+-5}{\milli\tesla}$ and $\SI{540+-34}{\milli\tesla}$ were required to suppress superconductivity, which is in good correspondence with the magnetometry data.

\begin{figure}
\includegraphics{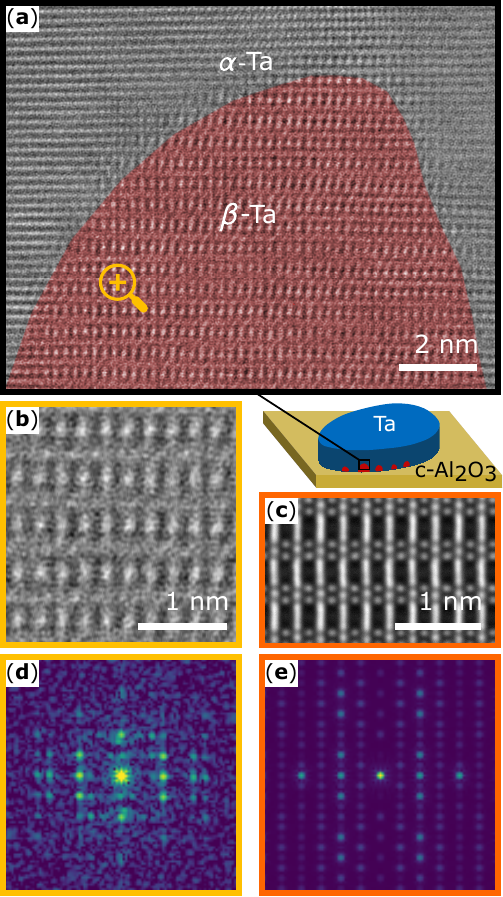}
\caption{Scanning transmission electron microscopy (STEM) images of $\alpha$-$\beta$ interface in the mixed-phase film. The schematic evokes the film structure from the Automated Crystal Orientation Mapping data in~\crefadd{fig: overview of films}{(b)} with scattered islands of $\beta$-phase close to Ta-sapphire interface. (a) High angle annular dark field (HAADF) STEM image shows a $\beta$-phase grain highlighted in red for better visibility. Its interface with the $\alpha$-phase is epitaxial. (b) and (c) compare the $\beta$-phase HAADF STEM image (yellow frame) with a simulated image (orange frame) at the same scale. (d) and (e) show fast Fourier transform of the images shown partially in (b) and (c), respectively and the width of the both images is $\SI{20.1}{\per\nano\m}$.}
\label{fig: TEM of mixed phase film}
\end{figure}

\cref{fig: TEM of mixed phase film} gives a deeper insight into the atomic-scale microstructure of the mixed-phase film. The $\beta$-phase forms a coherent interface with the $\alpha$-phase and furthermore confirms that the observed $\beta$-phase crystal structure corresponds to literature. The crystal unit cell of $\beta$-phase Ta is more complex than a primitive tetragonal unit cell~\cite{Moseley1973}, as shown in the bottom-mid panel of~\crefadd{fig: overview of films}{(b)}. From this, an overall more complex microstructure with increased amounts of high-angle grain boundaries potentially associated with a local suppression of superconductivity, could be expected. However, the $\beta$-phase forms a highly coherent interphase with the $\alpha$-phase~[\crefadd{fig: TEM of mixed phase film}{(a)}]. Additionally, we simulate the high angle annular dark field (HAADF) STEM image in~\crefadd{fig: TEM of mixed phase film}{(c)} (Crystallography data retrieved from the Materials Project for Ta (mp-42) from database version v2023.11.1.~\cite{materialsproject2013,Arakcheeva:sn0032}) and compare it with the measured HAADF image in~\crefadd{fig: TEM of mixed phase film}{(b)}. We observe a fair agreement between the experimental and the simulated structure in terms of observed lattice spacings. However, their fast Fourier transforms~[\crefadd{fig: TEM of mixed phase film}{(d,e)}] indicate differences in symmetry, which suggests small deviations from the theoretical crystal structure of the $\beta$-phase. Nevertheless, the images substantiate the claim that the dark areas in the ACOM data of the mixed-phase film in~\crefadd{fig: overview of films}{(b)} are closely related to the $\beta$-phase of Ta.

\begin{figure*}
    \centering
    \includegraphics[width=2\columnwidth]{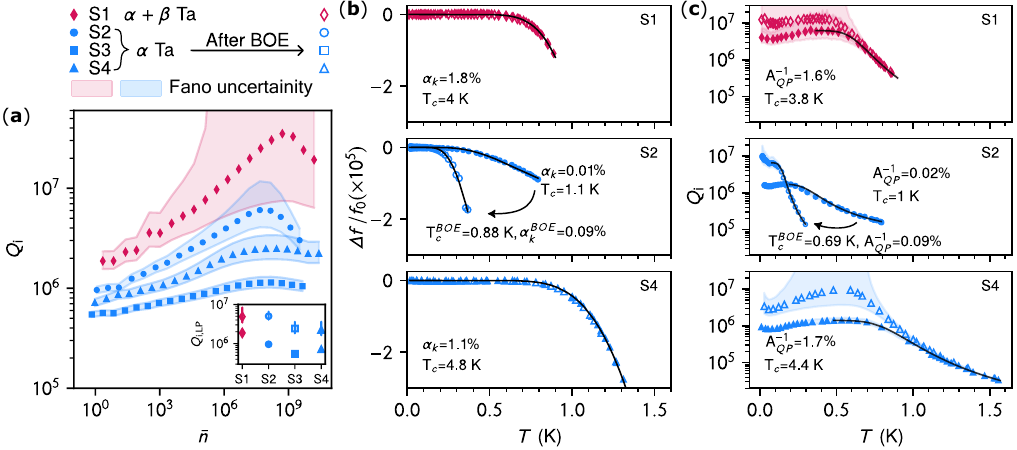}
    \caption{Microwave characterization of single- and mixed-phase Ta resonators before (filled markers) and after treating resonators in buffered-oxide etch (BOE, empty markers). The legend at the top left is common for all panels. The resonator design is shown in~\crefadd{fig: overview of films}{(c)}. (a) Internal quality factor $Q_\text{i}$ as a function of average photon number $\bar{n}$ with Fano uncertainty~\cite{Rieger2023Fano} shown as shaded interval and exceeds fitting uncertainty. The inset shows the improvement in $Q_\text{i, LP}$ ($Q_\text{i}$ at low power (LP)) for all four samples. (b) Relative frequency shift before and after treating resonators in BOE as a function of temperature, \(\Delta{f}/f_{0} = (f(T)-f_0)/f_0\), where $f_0$ is the resonant frequency at base temperature ($\approx\SI{10}{\milli\K}$). (c) $Q_\text{i}$ as a function of temperature before and after treating resonators in BOE. The black lines in (b) and (c) are fits to the simplified model given by~\cref{eq:MBfreq,eq:Q_qp} for the dataset before BOE treatment. For S2, we note a large change and therefore fit the dataset after BOE separately. The values for $T_\text{c}$ extracted from these fits are shown in the labels. In both (b) and (c) we keep the output power of the vector network analyzer constant, resulting in approximately $10^3$ average photons in the resonator at base temperature.} 
    \label{fig: microwave characterization}
\end{figure*}

We fabricated lumped element resonators (based on the design in Ref.~\cite{Winkel2020}) from both mixed- and single-phase films using e-beam (S1, S2) or optical (S3, S4) lithography (see SM).

The microwave measurements were done in a dilution cryostat (See SM) using a home-built copper 3D-waveguide sample holder~\cite{LukasG2018,Winkel2020}~[\crefadd{fig: overview of films}{(c)}] which provides a low loss environment for resonator characterization. The resonators are designed to have a high $Q_\text{c}$ of the order of $10^6$, which we achieve by fully closing the outer capacitor plate~[\crefadd{fig: overview of films}{(c)}]. This allows us to accurately determine internal losses with minimal uncertainty from the Fano effect~\cite{Rieger2023Fano}. 

In~\crefadd{fig: microwave characterization}{(a)} we characterize single- and mixed-phase Ta resonators by reporting $Q_\text{i}$ as a function of photon number $\bar{n}$~\cite{LukasG2018}. Surprisingly, the mixed-phase Ta resonator has the highest $Q_\text{i}$ of \num{2.0+-0.4d6} in the single photon regime. For single-phase Ta resonators, $Q_\text{i}$ ranges from \num{5d5} to $10^6$. All resonator $Q_\text{i}$ strongly increase with $\bar{n}$, which is associated with the presence of a saturable loss mechanism. However, in the regime $\bar{n}\gg 10^7$ we observe a reduction in $Q_\text{i}$ with $\bar{n}$, possibly associated with quasi-particle heating~\cite{McRae2020,Visser2014}. 

The inset plot in~\crefadd{fig: microwave characterization}{(a)} shows the improvement in $Q_\text{i, LP}$, defined as $Q_\text{i}$ at low-power (LP), after immersing the chips in BOE for $\SI{15}{\minute}$. We observe 3-6 fold improvement in $Q_\text{i, LP}$ irrespective of the film micro-structure (see SM for further evidence). Note that S1 and S2 resonators have comparable $Q_\text{i, LP}$ after BOE treatment. Furthermore, the $Q_\text{i, LP}$ deteriorates over weeks in air after BOE treatment and the improvement can be reproduced with successive BOE treatments (see SM). This points to a reversible loss mechanism. Possible candidates include etching and re-growth of the surface oxide or loading and unloading of Ta with hydrogen~\cite{Ni2020boe}. In the latter case hydrogen could potentially reduce loss from two-level systems~\cite{Muller2019Oct} by passivating dangling bonds. In addition, residual stress and strain in the film is affected by hydrogen content, which is another potential candidate for the dominating loss mechanism~\cite{Petersen2024stress-induced}. 

The temperature dependence of frequency shift and $Q_\text{i}$ for single- and mixed-phase Ta resonators is shown in~\crefadd{fig: microwave characterization}{(b,c)}, to detect possible contributions from any low-$T_\text{c}$ phases. We performed these measurements before and after BOE treatment for a selection of resonators. The frequency shift behaviour can be approximated by~\cite{Turneaure1991Oct}

\begin{equation}
    \frac{\Delta{f}(T)}{f_{0}} = -\frac{\alpha_\text{k}}{2}\sqrt{\frac{\pi\Delta_{0}}{2k_{B}T}}e^{\frac{-\Delta_{0}}{k_{B}T}},
\label{eq:MBfreq}    
\end{equation}

where $\alpha_\text{k}$ is the kinetic inductance fraction, $\Delta_{0} = 1.764~k_{B}T_\text{c}$ and $k_{B}$ is the Boltzmann constant. We use~\cref{eq:MBfreq} to fit the data in~\crefadd{fig: microwave characterization}{(b)}, keeping $\alpha_\text{k}$ and $T_\text{c}$ as free variables. Note that the two variables are correlated, which causes uncertainty in the fit result of a few percent. The $Q_\text{i}$ measurements~[\crefadd{fig: microwave characterization}{(c)}] exhibit an initial dip followed by an increase with increasing temperature, which  are associated with saturable loss observed in~\crefadd{fig: microwave characterization}{(a)}: For temperature sweep measurements at high power (not shown) the low temperature $Q_\text{i}$ is temperature independent. In the high temperature limit, $Q_\text{i}$ decreases due to the same quasi-particle excitations that cause the frequency shift in~\crefadd{fig: microwave characterization}{(b)}. This thermally activated loss mechanism is well described by the following equation~\cite{Crowley2023disentangling_losses_Ta,Gao_Jiansong} based on the Mattis-Bardeen model~\cite{Mattis1985}:

\begin{equation}
    Q_\text{QP}(T) = A_\text{QP} \frac{e^{\frac{\Delta_{0}}{k_{B}T}}}{\sinh(\frac{\hbar\omega}{2k_{B}T})K_{0}(\frac{\hbar\omega}{2k_{B}T})},
    \label{eq:Q_qp}
\end{equation}

where $A_\text{QP}$ is a constant proportional to $1/\alpha_\text{k}$, $\hbar$ is the reduced Planck constant, $\omega$ is the angular frequency, and $K_{0}$ is the zeroth order modified Bessel function of the second kind. We show fits to this model in~\crefadd{fig: microwave characterization}{(c)} approximating the low temperature loss with a constant. $A_\text{QP}$ and $T_\text{c}$ are taken as free parameters determined in the fit and are shown in~\crefadd{fig: microwave characterization}{(c)}. 

For S3 (not shown, same behaviour as S4) and S4, the extracted values for $T_\text{c}$, are in agreement with the gap and critical temperature measurements shown in~\cref{fig:Tc_and_delta} and are approximately consistent with those for bulk Ta of $\SI{4.4}{\K}$~\cite{Read1965Aug}. Similarly, for these two samples, the extracted values for $\alpha_\text{k}$ from the fit to~\cref{eq:MBfreq} are comparable to those extracted from the relative frequency shift between the measurement and microwave simulation $\alpha_{\rm k}^{\rm \Delta f_0} = 1-(f_0/ f_0^{\rm sim})^2\simeq 4\%$, where the simulation for $f_0^{\rm sim}$ does not take kinetic inductance into account (see SM). In the case of S1, the fits only result in a slightly lower $T_\text{c}$, showing that the $\beta$-phase grains are well proximitized~\cite{Catelani2022}. This is consistent with the atomically coherent interface between $\alpha-\beta$ phase observed in TEM~[\crefadd{fig: TEM of mixed phase film}{(a)}] and 4-probe DC transport measurements (see SM). 

In contrast, the $T_\text{c}$ values for S2 are significantly depressed (see also Ref.~\cite{Crowley2023disentangling_losses_Ta}). BOE treatment reduces the fitted $T_\text{c}$ for S2 further. Elemental analysis using time-of-flight secondary ion mass spectrometry did not reveal the presence of magnetic impurities that could explain the lower $T_\text{c}$ of S2 (see SM). Note that the values for $\alpha_\text{k}$ are 1-2 orders of magnitude lower compared to those in other samples and more importantly to $\alpha_\text{k}^{\Delta\text{f}_0}$ for that resonator. This large deviation suggests that only a small fraction of the order of $1-10\%$ of the resonator has a low $T_\text{c}$. To account for this, we propose the existence of a $\beta$-phase that remained undetected during film characterization due to its inhomogeneous distribution. Since $Q_\text{i, LP}$ of S2 was also high among the measured resonators, this hypothesis would align with the benefits of $\beta$-Ta in reducing microwave losses.

In summary, we have demonstrated that low-loss superconducting resonators can be fabricated from Ta films with a significant amount of $\beta$-phase defects at the metal-substrate interface. These mixed-phase Ta resonators with a $\SI{50}{\micro\meter}$ capacitor gap and 10\% volume fraction of $\beta$-phase achieved internal quality factors of \num{5.0+-2.5d6} in the single-photon regime. 
For resonators fabricated from high quality single-phase Ta films, single-photon quality factors were inferior in general. For one of these $\alpha$-phase resonators similarly high internal quality factor was measured. However, the temperature dependence of resonance frequencies and quality factors suggest that low-$T_\text{c}$ defects are present in the film, possibly associated with trace amounts of the $\beta$-phase not detectable in XRD analysis, or introduced by sample fabrication processes. The correlation between $\beta$-phase concentration and $Q_\text{i}$ is more complex than anticipated. We conclude that a small amount of $\beta$-phase in the film can be beneficial, enhancing critical magnetic fields and its concentration might even be a possible optimization parameter leading to improved coherence in Ta based superconducting circuits.

\section*{ACKNOWLEDGEMENTS}
The authors are grateful to L. Radtke and S. Dehm for technical support and to our collaborators from University of Innsbruck, Austria for providing tantalum films. Facilities use is supported by the Karlsruhe Institute of Technology (KIT) Nanostructure Service Laboratory (NSL) and the Karlsruhe Nano Micro Facility (KNMFi). We acknowledge qKit for providing a convenient measurement software framework.

This project has received funding from the Helmholtz Association, the European Union’s Horizon 2020 research and innovation programme under the Marie Skłodowska-Curie grant agreement number 847471, the Federal Ministry of Education and Research (Projects QSolid (FKZ:13N16151, FKZ 13N16159) and GeQCoS (FKZ: 13N15683)) and the Baden-Württemberg Stiftung under project QT-10 (QEDHiNet). In addition, J.A. and H.D.  acknowledge funding by the Joint-Lab Virtual Materials Design initiative and the Deutsche Forschungsgemeinschaft (DFG, German Research Foundation) under Germany's Excellence Strategy via the Excellence Cluster “3D Matter Made to Order” (EXC-2082/1-390761711), which has also been supported by the Carl Zeiss Foundation through the “Carl-Zeiss-Foundation Focus$@$HEiKA”, by the State of Baden-Württemberg and KIT.

\section*{author declarations}

\section*{Conflict of interest}
The authors have no conflicts to disclose.

\section*{Author Contributions}
\textbf{Ritika~Dhundhwal}:  Investigation (lead); Data curation (equal); Formal analysis (equal); Methodology (equal); Validation (equal); Visualization (equal); Writing - original draft (lead); Writing - review \& editing (equal).
\textbf{Haoran~Duan}: Investigation (equal); Data curation (equal); Formal analysis (equal); Writing - review \& editing (equal).
\textbf{Lucas~Brauch}: Investigation (equal); Data curation (equal); Formal analysis (equal); Writing - review \& editing (equal).
\textbf{Soroush~Arabi}: Investigation (equal); Data curation (equal); Formal analysis (equal); Writing - review \& editing (equal).
\textbf{Dirk~Fuchs}: Investigation (equal); Data curation (equal); Formal analysis (equal); Writing - review \& editing (equal).
\textbf{Amir-Abbas~Haghighirad}: Conceptualization (equal) ; Writing - review \& editing (equal).
\textbf{Alexander~Welle}: Investigation (equal); Data curation (equal); Formal analysis (equal); Writing - review \& editing (equal).
\textbf{Florentine~Scharwaechter}: Investigation (equal); Data curation (equal); Formal analysis (equal); Writing - review \& editing (equal).
\textbf{Sudip~Pal}: Investigation (equal); Data curation (equal); Formal analysis (equal); Writing - review \& editing (equal).
\textbf{Marc~Scheffler}: Conceptualization (equal) ;  Methodology (equal); Writing - review \& editing (equal).
\textbf{Jos\'e~Palomo}: Investigation (equal); Data curation (equal); Writing - review \& editing (equal).
\textbf{Zaki~Leghtas}: Conceptualization (equal) ;  Methodology (equal); Writing - review \& editing (equal).
\textbf{Anil~Murani}: Conceptualization (equal);  Methodology (equal); Writing - review \& editing (equal).
\textbf{Horst~Hahn}: Resources (equal); Funding acquisition (equal); Writing - review \& editing (equal)
\textbf{Jasmin~Aghassi-Hagmann}: Conceptualization (equal);  Methodology (equal); Writing - review \& editing (equal). 
\textbf{Christian~K\"ubel}: Conceptualization (equal);  Methodology (equal); Writing - review \& editing (equal). 
\textbf{Wulf~Wulfhekel}: Conceptualization (equal);  Methodology (equal); Writing - review \& editing (equal). 
\textbf{Ioan~M.~Pop}: Conceptualization (lead); Methodology (equal); Writing - review \& editing (equal); Resources (equal); Project administration (equal); Funding acquisition (equal).
\textbf{Thomas~Reisinger}: Conceptualization (equal); Methodology (equal); Writing - review \& editing (equal); Resources (equal); Project administration (lead); Funding acquisition (lead); Supervision (lead).

\section*{Data Availability}
Data supporting the findings of this study are available from the corresponding author upon reasonable request.

\bibliography{aipsamp}

\providecommand{\noopsort}[1]{}\providecommand{\singleletter}[1]{#1}%
\begin{thebibliography}{44}%
\makeatletter
\providecommand \@ifxundefined [1]{%
 \@ifx{#1\undefined}
}%
\providecommand \@ifnum [1]{%
 \ifnum #1\expandafter \@firstoftwo
 \else \expandafter \@secondoftwo
 \fi
}%
\providecommand \@ifx [1]{%
 \ifx #1\expandafter \@firstoftwo
 \else \expandafter \@secondoftwo
 \fi
}%
\providecommand \natexlab [1]{#1}%
\providecommand \enquote  [1]{``#1''}%
\providecommand \bibnamefont  [1]{#1}%
\providecommand \bibfnamefont [1]{#1}%
\providecommand \citenamefont [1]{#1}%
\providecommand \href@noop [0]{\@secondoftwo}%
\providecommand \href [0]{\begingroup \@sanitize@url \@href}%
\providecommand \@href[1]{\@@startlink{#1}\@@href}%
\providecommand \@@href[1]{\endgroup#1\@@endlink}%
\providecommand \@sanitize@url [0]{\catcode `\\12\catcode `\$12\catcode `\&12\catcode `\#12\catcode `\^12\catcode `\_12\catcode `\%12\relax}%
\providecommand \@@startlink[1]{}%
\providecommand \@@endlink[0]{}%
\providecommand \url  [0]{\begingroup\@sanitize@url \@url }%
\providecommand \@url [1]{\endgroup\@href {#1}{\urlprefix }}%
\providecommand \urlprefix  [0]{URL }%
\providecommand \Eprint [0]{\href }%
\providecommand \doibase [0]{https://doi.org/}%
\providecommand \selectlanguage [0]{\@gobble}%
\providecommand \bibinfo  [0]{\@secondoftwo}%
\providecommand \bibfield  [0]{\@secondoftwo}%
\providecommand \translation [1]{[#1]}%
\providecommand \BibitemOpen [0]{}%
\providecommand \bibitemStop [0]{}%
\providecommand \bibitemNoStop [0]{.\EOS\space}%
\providecommand \EOS [0]{\spacefactor3000\relax}%
\providecommand \BibitemShut  [1]{\csname bibitem#1\endcsname}%
\let\auto@bib@innerbib\@empty
\bibitem [{\citenamefont {Blais}\ \emph {et~al.}(2021)\citenamefont {Blais}, \citenamefont {Grimsmo}, \citenamefont {Girvin},\ and\ \citenamefont {Wallraff}}]{Blais2021cqed}%
  \BibitemOpen
  \bibfield  {author} {\bibinfo {author} {\bibfnamefont {A.}~\bibnamefont {Blais}}, \bibinfo {author} {\bibfnamefont {A.~L.}\ \bibnamefont {Grimsmo}}, \bibinfo {author} {\bibfnamefont {S.~M.}\ \bibnamefont {Girvin}},\ and\ \bibinfo {author} {\bibfnamefont {A.}~\bibnamefont {Wallraff}},\ }\bibfield  {title} {\enquote {\bibinfo {title} {Circuit quantum electrodynamics},}\ }\href {https://doi.org/10.1103/RevModPhys.93.025005} {\bibfield  {journal} {\bibinfo  {journal} {Rev. Mod. Phys.}\ }\textbf {\bibinfo {volume} {93}},\ \bibinfo {pages} {025005} (\bibinfo {year} {2021})}\BibitemShut {NoStop}%
\bibitem [{\citenamefont {Krasnok}\ \emph {et~al.}(2024)\citenamefont {Krasnok}, \citenamefont {Dhakal}, \citenamefont {Fedorov}, \citenamefont {Frigola}, \citenamefont {Kelly},\ and\ \citenamefont {Kutsaev}}]{Krasnok_2024}%
  \BibitemOpen
  \bibfield  {author} {\bibinfo {author} {\bibfnamefont {A.}~\bibnamefont {Krasnok}}, \bibinfo {author} {\bibfnamefont {P.}~\bibnamefont {Dhakal}}, \bibinfo {author} {\bibfnamefont {A.}~\bibnamefont {Fedorov}}, \bibinfo {author} {\bibfnamefont {P.}~\bibnamefont {Frigola}}, \bibinfo {author} {\bibfnamefont {M.}~\bibnamefont {Kelly}},\ and\ \bibinfo {author} {\bibfnamefont {S.}~\bibnamefont {Kutsaev}},\ }\bibfield  {title} {\enquote {\bibinfo {title} {Superconducting microwave cavities and qubits for quantum information systems},}\ }\href {https://doi.org/10.1063/5.0155213} {\bibfield  {journal} {\bibinfo  {journal} {Applied Physics Reviews}\ }\textbf {\bibinfo {volume} {11}},\ \bibinfo {pages} {011302} (\bibinfo {year} {2024})}\BibitemShut {NoStop}%
\bibitem [{\citenamefont {Krantz}\ \emph {et~al.}(2019)\citenamefont {Krantz}, \citenamefont {Kjaergaard}, \citenamefont {Yan}, \citenamefont {Orlando}, \citenamefont {Gustavsson},\ and\ \citenamefont {Oliver}}]{Krantz2019Jun}%
  \BibitemOpen
  \bibfield  {author} {\bibinfo {author} {\bibfnamefont {P.}~\bibnamefont {Krantz}}, \bibinfo {author} {\bibfnamefont {M.}~\bibnamefont {Kjaergaard}}, \bibinfo {author} {\bibfnamefont {F.}~\bibnamefont {Yan}}, \bibinfo {author} {\bibfnamefont {T.~P.}\ \bibnamefont {Orlando}}, \bibinfo {author} {\bibfnamefont {S.}~\bibnamefont {Gustavsson}},\ and\ \bibinfo {author} {\bibfnamefont {W.~D.}\ \bibnamefont {Oliver}},\ }\bibfield  {title} {\enquote {\bibinfo {title} {{A quantum engineer's guide to superconducting qubits}},}\ }\href {https://doi.org/10.1063/1.5089550} {\bibfield  {journal} {\bibinfo  {journal} {Appl. Phys. Rev.}\ }\textbf {\bibinfo {volume} {6}} (\bibinfo {year} {2019}),\ 10.1063/1.5089550}\BibitemShut {NoStop}%
\bibitem [{\citenamefont {Arute}\ \emph {et~al.}(2019)\citenamefont {Arute}, \citenamefont {Arya}, \citenamefont {Babbush}, \citenamefont {Bacon}, \citenamefont {Bardin}, \citenamefont {Barends}, \citenamefont {Biswas}, \citenamefont {Boixo}, \citenamefont {Brandao}, \citenamefont {Buell}, \citenamefont {Burkett}, \citenamefont {Chen}, \citenamefont {Chen}, \citenamefont {Chiaro}, \citenamefont {Collins}, \citenamefont {Courtney}, \citenamefont {Dunsworth}, \citenamefont {Farhi}, \citenamefont {Foxen}, \citenamefont {Fowler}, \citenamefont {Gidney}, \citenamefont {Giustina}, \citenamefont {Graff}, \citenamefont {Guerin}, \citenamefont {Habegger}, \citenamefont {Harrigan}, \citenamefont {Hartmann}, \citenamefont {Ho}, \citenamefont {Hoffmann}, \citenamefont {Huang}, \citenamefont {Humble}, \citenamefont {Isakov}, \citenamefont {Jeffrey}, \citenamefont {Jiang}, \citenamefont {Kafri}, \citenamefont {Kechedzhi}, \citenamefont {Kelly}, \citenamefont {Knysh}, \citenamefont {Korotkov}, \citenamefont {Kostritsa},
  \citenamefont {Landhuis}, \citenamefont {Lindmark}, \citenamefont {Lucero}, \citenamefont {Lyakh}, \citenamefont {Mandrà}, \citenamefont {McClean}, \citenamefont {McEwen}, \citenamefont {Megrant}, \citenamefont {Mi}, \citenamefont {Michielsen}, \citenamefont {Mohseni}, \citenamefont {Mutus}, \citenamefont {Naaman}, \citenamefont {Neeley}, \citenamefont {Neill}, \citenamefont {Niu}, \citenamefont {Ostby}, \citenamefont {Petukhov}, \citenamefont {Platt}, \citenamefont {Quintana}, \citenamefont {Rieffel}, \citenamefont {Roushan}, \citenamefont {Rubin}, \citenamefont {Sank}, \citenamefont {Satzinger}, \citenamefont {Smelyanskiy}, \citenamefont {Sung}, \citenamefont {Trevithick}, \citenamefont {Vainsencher}, \citenamefont {Villalonga}, \citenamefont {White}, \citenamefont {Yao}, \citenamefont {Yeh}, \citenamefont {Zalcman}, \citenamefont {Neven},\ and\ \citenamefont {Martin}}]{Quantumsupremacy2019}%
  \BibitemOpen
  \bibfield  {author} {\bibinfo {author} {\bibfnamefont {F.}~\bibnamefont {Arute}}, \bibinfo {author} {\bibfnamefont {K.}~\bibnamefont {Arya}}, \bibinfo {author} {\bibfnamefont {R.}~\bibnamefont {Babbush}}, \bibinfo {author} {\bibfnamefont {D.}~\bibnamefont {Bacon}}, \bibinfo {author} {\bibfnamefont {J.~C.}\ \bibnamefont {Bardin}}, \bibinfo {author} {\bibfnamefont {R.}~\bibnamefont {Barends}}, \bibinfo {author} {\bibfnamefont {R.}~\bibnamefont {Biswas}}, \bibinfo {author} {\bibfnamefont {S.}~\bibnamefont {Boixo}}, \bibinfo {author} {\bibfnamefont {F.~G. S.~L.}\ \bibnamefont {Brandao}}, \bibinfo {author} {\bibfnamefont {D.~A.}\ \bibnamefont {Buell}}, \bibinfo {author} {\bibfnamefont {B.}~\bibnamefont {Burkett}}, \bibinfo {author} {\bibfnamefont {Y.}~\bibnamefont {Chen}}, \bibinfo {author} {\bibfnamefont {Z.}~\bibnamefont {Chen}}, \bibinfo {author} {\bibfnamefont {B.}~\bibnamefont {Chiaro}}, \bibinfo {author} {\bibfnamefont {R.}~\bibnamefont {Collins}}, \bibinfo {author} {\bibfnamefont {W.}~\bibnamefont
  {Courtney}}, \bibinfo {author} {\bibfnamefont {A.}~\bibnamefont {Dunsworth}}, \bibinfo {author} {\bibfnamefont {E.}~\bibnamefont {Farhi}}, \bibinfo {author} {\bibfnamefont {B.}~\bibnamefont {Foxen}}, \bibinfo {author} {\bibfnamefont {A.}~\bibnamefont {Fowler}}, \bibinfo {author} {\bibfnamefont {C.}~\bibnamefont {Gidney}}, \bibinfo {author} {\bibfnamefont {M.}~\bibnamefont {Giustina}}, \bibinfo {author} {\bibfnamefont {R.}~\bibnamefont {Graff}}, \bibinfo {author} {\bibfnamefont {K.}~\bibnamefont {Guerin}}, \bibinfo {author} {\bibfnamefont {S.}~\bibnamefont {Habegger}}, \bibinfo {author} {\bibfnamefont {M.~P.}\ \bibnamefont {Harrigan}}, \bibinfo {author} {\bibfnamefont {M.~J.}\ \bibnamefont {Hartmann}}, \bibinfo {author} {\bibfnamefont {A.}~\bibnamefont {Ho}}, \bibinfo {author} {\bibfnamefont {M.}~\bibnamefont {Hoffmann}}, \bibinfo {author} {\bibfnamefont {T.}~\bibnamefont {Huang}}, \bibinfo {author} {\bibfnamefont {T.~S.}\ \bibnamefont {Humble}}, \bibinfo {author} {\bibfnamefont {S.~V.}\ \bibnamefont
  {Isakov}}, \bibinfo {author} {\bibfnamefont {E.}~\bibnamefont {Jeffrey}}, \bibinfo {author} {\bibfnamefont {Z.}~\bibnamefont {Jiang}}, \bibinfo {author} {\bibfnamefont {D.}~\bibnamefont {Kafri}}, \bibinfo {author} {\bibfnamefont {K.}~\bibnamefont {Kechedzhi}}, \bibinfo {author} {\bibfnamefont {P.~V.}\ \bibnamefont {Kelly}, \bibfnamefont {and~Klimov}}, \bibinfo {author} {\bibfnamefont {S.}~\bibnamefont {Knysh}}, \bibinfo {author} {\bibfnamefont {A.}~\bibnamefont {Korotkov}}, \bibinfo {author} {\bibfnamefont {F.}~\bibnamefont {Kostritsa}}, \bibinfo {author} {\bibfnamefont {D.}~\bibnamefont {Landhuis}}, \bibinfo {author} {\bibfnamefont {M.}~\bibnamefont {Lindmark}}, \bibinfo {author} {\bibfnamefont {E.}~\bibnamefont {Lucero}}, \bibinfo {author} {\bibfnamefont {D.}~\bibnamefont {Lyakh}}, \bibinfo {author} {\bibfnamefont {S.}~\bibnamefont {Mandrà}}, \bibinfo {author} {\bibfnamefont {J.~R.}\ \bibnamefont {McClean}}, \bibinfo {author} {\bibfnamefont {M.}~\bibnamefont {McEwen}}, \bibinfo {author} {\bibfnamefont
  {A.}~\bibnamefont {Megrant}}, \bibinfo {author} {\bibfnamefont {X.}~\bibnamefont {Mi}}, \bibinfo {author} {\bibfnamefont {K.}~\bibnamefont {Michielsen}}, \bibinfo {author} {\bibfnamefont {M.}~\bibnamefont {Mohseni}}, \bibinfo {author} {\bibfnamefont {J.}~\bibnamefont {Mutus}}, \bibinfo {author} {\bibfnamefont {O.}~\bibnamefont {Naaman}}, \bibinfo {author} {\bibfnamefont {M.}~\bibnamefont {Neeley}}, \bibinfo {author} {\bibfnamefont {C.}~\bibnamefont {Neill}}, \bibinfo {author} {\bibfnamefont {M.~Y.}\ \bibnamefont {Niu}}, \bibinfo {author} {\bibfnamefont {E.}~\bibnamefont {Ostby}}, \bibinfo {author} {\bibfnamefont {A.}~\bibnamefont {Petukhov}}, \bibinfo {author} {\bibfnamefont {J.~C.}\ \bibnamefont {Platt}}, \bibinfo {author} {\bibfnamefont {C.}~\bibnamefont {Quintana}}, \bibinfo {author} {\bibfnamefont {E.~G.}\ \bibnamefont {Rieffel}}, \bibinfo {author} {\bibfnamefont {P.}~\bibnamefont {Roushan}}, \bibinfo {author} {\bibfnamefont {N.~C.}\ \bibnamefont {Rubin}}, \bibinfo {author} {\bibfnamefont
  {D.}~\bibnamefont {Sank}}, \bibinfo {author} {\bibfnamefont {K.~J.}\ \bibnamefont {Satzinger}}, \bibinfo {author} {\bibfnamefont {V.}~\bibnamefont {Smelyanskiy}}, \bibinfo {author} {\bibfnamefont {K.~J.}\ \bibnamefont {Sung}}, \bibinfo {author} {\bibfnamefont {M.~D.}\ \bibnamefont {Trevithick}}, \bibinfo {author} {\bibfnamefont {A.}~\bibnamefont {Vainsencher}}, \bibinfo {author} {\bibfnamefont {B.}~\bibnamefont {Villalonga}}, \bibinfo {author} {\bibfnamefont {T.}~\bibnamefont {White}}, \bibinfo {author} {\bibfnamefont {Z.~J.}\ \bibnamefont {Yao}}, \bibinfo {author} {\bibfnamefont {P.}~\bibnamefont {Yeh}}, \bibinfo {author} {\bibfnamefont {A.}~\bibnamefont {Zalcman}}, \bibinfo {author} {\bibfnamefont {H.}~\bibnamefont {Neven}},\ and\ \bibinfo {author} {\bibfnamefont {J.~M.}\ \bibnamefont {Martin}},\ }\bibfield  {title} {\enquote {\bibinfo {title} {Quantum supremacy using a programmable superconducting processor},}\ }\href {https://doi.org/10.1038/s41586-019-1666-5} {\bibfield  {journal} {\bibinfo  {journal}
  {Nature}\ }\textbf {\bibinfo {volume} {574}},\ \bibinfo {pages} {505--510} (\bibinfo {year} {2019})}\BibitemShut {NoStop}%
\bibitem [{\citenamefont {Sivak}\ \emph {et~al.}(2023)\citenamefont {Sivak}, \citenamefont {Eickbusch}, \citenamefont {Royer}, \citenamefont {Singh}, \citenamefont {Tsioutsios}, \citenamefont {Ganjam}, \citenamefont {Miano}, \citenamefont {Brock}, \citenamefont {Ding}, \citenamefont {Frunzio}, \citenamefont {Girvin}, \citenamefont {Schoelkopf},\ and\ \citenamefont {Devoret}}]{sivak2023}%
  \BibitemOpen
  \bibfield  {author} {\bibinfo {author} {\bibfnamefont {V.~V.}\ \bibnamefont {Sivak}}, \bibinfo {author} {\bibfnamefont {A.}~\bibnamefont {Eickbusch}}, \bibinfo {author} {\bibfnamefont {B.}~\bibnamefont {Royer}}, \bibinfo {author} {\bibfnamefont {S.}~\bibnamefont {Singh}}, \bibinfo {author} {\bibfnamefont {I.}~\bibnamefont {Tsioutsios}}, \bibinfo {author} {\bibfnamefont {S.}~\bibnamefont {Ganjam}}, \bibinfo {author} {\bibfnamefont {A.}~\bibnamefont {Miano}}, \bibinfo {author} {\bibfnamefont {B.~L.}\ \bibnamefont {Brock}}, \bibinfo {author} {\bibfnamefont {A.~Z.}\ \bibnamefont {Ding}}, \bibinfo {author} {\bibfnamefont {L.}~\bibnamefont {Frunzio}}, \bibinfo {author} {\bibfnamefont {S.~M.}\ \bibnamefont {Girvin}}, \bibinfo {author} {\bibfnamefont {R.~J.}\ \bibnamefont {Schoelkopf}},\ and\ \bibinfo {author} {\bibfnamefont {M.~H.}\ \bibnamefont {Devoret}},\ }\bibfield  {title} {\enquote {\bibinfo {title} {Real-time quantum error correction beyond break-even},}\ }\href {https://doi.org/10.1038/s41586-023-05782-6}
  {\bibfield  {journal} {\bibinfo  {journal} {Nature}\ }\textbf {\bibinfo {volume} {616}},\ \bibinfo {pages} {50–55} (\bibinfo {year} {2023})}\BibitemShut {NoStop}%
\bibitem [{\citenamefont {Aumentado}(2020)}]{Jose2020_JPA}%
  \BibitemOpen
  \bibfield  {author} {\bibinfo {author} {\bibfnamefont {J.}~\bibnamefont {Aumentado}},\ }\bibfield  {title} {\enquote {\bibinfo {title} {Superconducting parametric amplifiers: The state of the art in josephson parametric amplifiers},}\ }\href {https://doi.org/10.1109/MMM.2020.2993476} {\bibfield  {journal} {\bibinfo  {journal} {IEEE Microwave Magazine}\ }\textbf {\bibinfo {volume} {21}},\ \bibinfo {pages} {45--59} (\bibinfo {year} {2020})}\BibitemShut {NoStop}%
\bibitem [{\citenamefont {Siddiqi}(0 01)}]{siddiqi_engineering_2021}%
  \BibitemOpen
  \bibfield  {author} {\bibinfo {author} {\bibfnamefont {I.}~\bibnamefont {Siddiqi}},\ }\bibfield  {title} {\enquote {\bibinfo {title} {Engineering high-coherence superconducting qubits},}\ }\href {https://doi.org/10.1038/s41578-021-00370-4} {\bibfield  {journal} {\bibinfo  {journal} {Nature Reviews Materials}\ }\textbf {\bibinfo {volume} {6}},\ \bibinfo {pages} {875--891} (\bibinfo {year} {2021-10-01})}\BibitemShut {NoStop}%
\bibitem [{\citenamefont {McRae}\ \emph {et~al.}(2020)\citenamefont {McRae}, \citenamefont {Wang}, \citenamefont {Gao}, \citenamefont {Vissers}, \citenamefont {Brecht}, \citenamefont {Dunsworth}, \citenamefont {Pappas},\ and\ \citenamefont {Mutus}}]{McRae2020}%
  \BibitemOpen
  \bibfield  {author} {\bibinfo {author} {\bibfnamefont {C.~R.~H.}\ \bibnamefont {McRae}}, \bibinfo {author} {\bibfnamefont {H.}~\bibnamefont {Wang}}, \bibinfo {author} {\bibfnamefont {J.}~\bibnamefont {Gao}}, \bibinfo {author} {\bibfnamefont {M.~R.}\ \bibnamefont {Vissers}}, \bibinfo {author} {\bibfnamefont {T.}~\bibnamefont {Brecht}}, \bibinfo {author} {\bibfnamefont {A.}~\bibnamefont {Dunsworth}}, \bibinfo {author} {\bibfnamefont {D.~P.}\ \bibnamefont {Pappas}},\ and\ \bibinfo {author} {\bibfnamefont {J.}~\bibnamefont {Mutus}},\ }\bibfield  {title} {\enquote {\bibinfo {title} {{Materials loss measurements using superconducting microwave resonators}},}\ }\href {https://doi.org/10.1063/5.0017378} {\bibfield  {journal} {\bibinfo  {journal} {Review of Scientific Instruments}\ }\textbf {\bibinfo {volume} {91}},\ \bibinfo {pages} {091101} (\bibinfo {year} {2020})}\BibitemShut {NoStop}%
\bibitem [{\citenamefont {de~Leon}\ \emph {et~al.}(2021)\citenamefont {de~Leon}, \citenamefont {Itoh}, \citenamefont {Kim}, \citenamefont {Mehta}, \citenamefont {Northup}, \citenamefont {Paik}, \citenamefont {Palmer}, \citenamefont {Samarth}, \citenamefont {Sangtawesin},\ and\ \citenamefont {Steuerman}}]{deLeon2021Apr}%
  \BibitemOpen
  \bibfield  {author} {\bibinfo {author} {\bibfnamefont {N.~P.}\ \bibnamefont {de~Leon}}, \bibinfo {author} {\bibfnamefont {K.~M.}\ \bibnamefont {Itoh}}, \bibinfo {author} {\bibfnamefont {D.}~\bibnamefont {Kim}}, \bibinfo {author} {\bibfnamefont {K.~K.}\ \bibnamefont {Mehta}}, \bibinfo {author} {\bibfnamefont {T.~E.}\ \bibnamefont {Northup}}, \bibinfo {author} {\bibfnamefont {H.}~\bibnamefont {Paik}}, \bibinfo {author} {\bibfnamefont {B.~S.}\ \bibnamefont {Palmer}}, \bibinfo {author} {\bibfnamefont {N.}~\bibnamefont {Samarth}}, \bibinfo {author} {\bibfnamefont {S.}~\bibnamefont {Sangtawesin}},\ and\ \bibinfo {author} {\bibfnamefont {D.~W.}\ \bibnamefont {Steuerman}},\ }\bibfield  {title} {\enquote {\bibinfo {title} {{Materials challenges and opportunities for quantum computing hardware}},}\ }\href {https://doi.org/10.1126/science.abb2823} {\bibfield  {journal} {\bibinfo  {journal} {Science}\ }\textbf {\bibinfo {volume} {372}} (\bibinfo {year} {2021}),\ 10.1126/science.abb2823}\BibitemShut {NoStop}%
\bibitem [{\citenamefont {Chayanun}\ \emph {et~al.}(2024)\citenamefont {Chayanun}, \citenamefont {Biznárová}, \citenamefont {Zeng}, \citenamefont {Malmberg}, \citenamefont {Nylander}, \citenamefont {Osman}, \citenamefont {Rommel}, \citenamefont {Tam}, \citenamefont {Olsson}, \citenamefont {Delsing}, \citenamefont {Yurgens}, \citenamefont {Bylander},\ and\ \citenamefont {Fadavi~Roudsari}}]{Chayanun2024_aluminum-on-silicon}%
  \BibitemOpen
  \bibfield  {author} {\bibinfo {author} {\bibfnamefont {L.}~\bibnamefont {Chayanun}}, \bibinfo {author} {\bibfnamefont {J.}~\bibnamefont {Biznárová}}, \bibinfo {author} {\bibfnamefont {L.}~\bibnamefont {Zeng}}, \bibinfo {author} {\bibfnamefont {P.}~\bibnamefont {Malmberg}}, \bibinfo {author} {\bibfnamefont {A.}~\bibnamefont {Nylander}}, \bibinfo {author} {\bibfnamefont {A.}~\bibnamefont {Osman}}, \bibinfo {author} {\bibfnamefont {M.}~\bibnamefont {Rommel}}, \bibinfo {author} {\bibfnamefont {P.~L.}\ \bibnamefont {Tam}}, \bibinfo {author} {\bibfnamefont {E.}~\bibnamefont {Olsson}}, \bibinfo {author} {\bibfnamefont {P.}~\bibnamefont {Delsing}}, \bibinfo {author} {\bibfnamefont {A.}~\bibnamefont {Yurgens}}, \bibinfo {author} {\bibfnamefont {J.}~\bibnamefont {Bylander}},\ and\ \bibinfo {author} {\bibfnamefont {A.}~\bibnamefont {Fadavi~Roudsari}},\ }\bibfield  {title} {\enquote {\bibinfo {title} {Characterization of process-related interfacial dielectric loss in aluminum-on-silicon by resonator microwave
  measurements, materials analysis, and imaging},}\ }\href {https://doi.org/10.1063/5.0208140} {\bibfield  {journal} {\bibinfo  {journal} {APL Quantum}\ }\textbf {\bibinfo {volume} {1}},\ \bibinfo {pages} {026115} (\bibinfo {year} {2024})}\BibitemShut {NoStop}%
\bibitem [{\citenamefont {Tuokkola}\ \emph {et~al.}(2024)\citenamefont {Tuokkola}, \citenamefont {Sunada}, \citenamefont {Kivijärvi}, \citenamefont {Albanese}, \citenamefont {Grönberg}, \citenamefont {Kaikkonen}, \citenamefont {Vesterinen}, \citenamefont {Govenius},\ and\ \citenamefont {Möttönen}}]{tuokkola2024methodsachievenearmillisecondenergy}%
  \BibitemOpen
  \bibfield  {author} {\bibinfo {author} {\bibfnamefont {M.}~\bibnamefont {Tuokkola}}, \bibinfo {author} {\bibfnamefont {Y.}~\bibnamefont {Sunada}}, \bibinfo {author} {\bibfnamefont {H.}~\bibnamefont {Kivijärvi}}, \bibinfo {author} {\bibfnamefont {J.}~\bibnamefont {Albanese}}, \bibinfo {author} {\bibfnamefont {L.}~\bibnamefont {Grönberg}}, \bibinfo {author} {\bibfnamefont {J.-P.}\ \bibnamefont {Kaikkonen}}, \bibinfo {author} {\bibfnamefont {V.}~\bibnamefont {Vesterinen}}, \bibinfo {author} {\bibfnamefont {J.}~\bibnamefont {Govenius}},\ and\ \bibinfo {author} {\bibfnamefont {M.}~\bibnamefont {Möttönen}},\ }\href {https://arxiv.org/abs/2407.18778} {\enquote {\bibinfo {title} {Methods to achieve near-millisecond energy relaxation and dephasing times for a superconducting transmon qubit},}\ } (\bibinfo {year} {2024}),\ \Eprint {https://arxiv.org/abs/2407.18778} {arXiv:2407.18778 [quant-ph]} \BibitemShut {NoStop}%
\bibitem [{\citenamefont {Place}\ \emph {et~al.}(2021)\citenamefont {Place}, \citenamefont {Rodgers}, \citenamefont {Mundada}, \citenamefont {Smitham}, \citenamefont {Fitzpatrick}, \citenamefont {Leng}, \citenamefont {Premkumar}, \citenamefont {Bryon}, \citenamefont {Vrajitoarea}, \citenamefont {Sussman} \emph {et~al.}}]{place2021}%
  \BibitemOpen
  \bibfield  {author} {\bibinfo {author} {\bibfnamefont {A.~P.}\ \bibnamefont {Place}}, \bibinfo {author} {\bibfnamefont {L.~V.}\ \bibnamefont {Rodgers}}, \bibinfo {author} {\bibfnamefont {P.}~\bibnamefont {Mundada}}, \bibinfo {author} {\bibfnamefont {B.~M.}\ \bibnamefont {Smitham}}, \bibinfo {author} {\bibfnamefont {M.}~\bibnamefont {Fitzpatrick}}, \bibinfo {author} {\bibfnamefont {Z.}~\bibnamefont {Leng}}, \bibinfo {author} {\bibfnamefont {A.}~\bibnamefont {Premkumar}}, \bibinfo {author} {\bibfnamefont {J.}~\bibnamefont {Bryon}}, \bibinfo {author} {\bibfnamefont {A.}~\bibnamefont {Vrajitoarea}}, \bibinfo {author} {\bibfnamefont {S.}~\bibnamefont {Sussman}}, \emph {et~al.},\ }\bibfield  {title} {\enquote {\bibinfo {title} {New material platform for superconducting transmon qubits with coherence times exceeding 0.3 milliseconds},}\ }\href {https://doi.org/10.1038/s41586-019-1666-5} {\bibfield  {journal} {\bibinfo  {journal} {Nature communications}\ }\textbf {\bibinfo {volume} {12}},\ \bibinfo {pages} {1779}
  (\bibinfo {year} {2021})}\BibitemShut {NoStop}%
\bibitem [{\citenamefont {Wang}\ \emph {et~al.}(2022)\citenamefont {Wang}, \citenamefont {Li}, \citenamefont {Xu}, \citenamefont {Li}, \citenamefont {Wang}, \citenamefont {Yang}, \citenamefont {Mi}, \citenamefont {Liang}, \citenamefont {Su}, \citenamefont {Yang} \emph {et~al.}}]{wang2022}%
  \BibitemOpen
  \bibfield  {author} {\bibinfo {author} {\bibfnamefont {C.}~\bibnamefont {Wang}}, \bibinfo {author} {\bibfnamefont {X.}~\bibnamefont {Li}}, \bibinfo {author} {\bibfnamefont {H.}~\bibnamefont {Xu}}, \bibinfo {author} {\bibfnamefont {Z.}~\bibnamefont {Li}}, \bibinfo {author} {\bibfnamefont {J.}~\bibnamefont {Wang}}, \bibinfo {author} {\bibfnamefont {Z.}~\bibnamefont {Yang}}, \bibinfo {author} {\bibfnamefont {Z.}~\bibnamefont {Mi}}, \bibinfo {author} {\bibfnamefont {X.}~\bibnamefont {Liang}}, \bibinfo {author} {\bibfnamefont {T.}~\bibnamefont {Su}}, \bibinfo {author} {\bibfnamefont {C.}~\bibnamefont {Yang}}, \emph {et~al.},\ }\bibfield  {title} {\enquote {\bibinfo {title} {Towards practical quantum computers: Transmon qubit with a lifetime approaching 0.5 milliseconds},}\ }\href {https://doi.org/10.1038/s41534-021-00510-2} {\bibfield  {journal} {\bibinfo  {journal} {npj Quantum Information}\ }\textbf {\bibinfo {volume} {8}},\ \bibinfo {pages} {3} (\bibinfo {year} {2022})}\BibitemShut {NoStop}%
\bibitem [{\citenamefont {Anferov}\ \emph {et~al.}(2024)\citenamefont {Anferov}, \citenamefont {Lee}, \citenamefont {Zhao}, \citenamefont {Simon},\ and\ \citenamefont {Schuster}}]{Anferov2024nb.trilayer.junction}%
  \BibitemOpen
  \bibfield  {author} {\bibinfo {author} {\bibfnamefont {A.}~\bibnamefont {Anferov}}, \bibinfo {author} {\bibfnamefont {K.-H.}\ \bibnamefont {Lee}}, \bibinfo {author} {\bibfnamefont {F.}~\bibnamefont {Zhao}}, \bibinfo {author} {\bibfnamefont {J.}~\bibnamefont {Simon}},\ and\ \bibinfo {author} {\bibfnamefont {D.~I.}\ \bibnamefont {Schuster}},\ }\bibfield  {title} {\enquote {\bibinfo {title} {Improved coherence in optically defined niobium trilayer-junction qubits},}\ }\href {https://doi.org/10.1103/PhysRevApplied.21.024047} {\bibfield  {journal} {\bibinfo  {journal} {Phys. Rev. Appl.}\ }\textbf {\bibinfo {volume} {21}},\ \bibinfo {pages} {024047} (\bibinfo {year} {2024})}\BibitemShut {NoStop}%
\bibitem [{\citenamefont {Chang}\ \emph {et~al.}(2013)\citenamefont {Chang}, \citenamefont {Vissers}, \citenamefont {Córcoles}, \citenamefont {Sandberg}, \citenamefont {Gao}, \citenamefont {Abraham}, \citenamefont {Chow}, \citenamefont {Gambetta}, \citenamefont {Beth~Rothwell}, \citenamefont {Keefe}, \citenamefont {Steffen},\ and\ \citenamefont {Pappas}}]{Chang2013}%
  \BibitemOpen
  \bibfield  {author} {\bibinfo {author} {\bibfnamefont {J.~B.}\ \bibnamefont {Chang}}, \bibinfo {author} {\bibfnamefont {M.~R.}\ \bibnamefont {Vissers}}, \bibinfo {author} {\bibfnamefont {A.~D.}\ \bibnamefont {Córcoles}}, \bibinfo {author} {\bibfnamefont {M.}~\bibnamefont {Sandberg}}, \bibinfo {author} {\bibfnamefont {J.}~\bibnamefont {Gao}}, \bibinfo {author} {\bibfnamefont {D.~W.}\ \bibnamefont {Abraham}}, \bibinfo {author} {\bibfnamefont {J.~M.}\ \bibnamefont {Chow}}, \bibinfo {author} {\bibfnamefont {J.~M.}\ \bibnamefont {Gambetta}}, \bibinfo {author} {\bibfnamefont {M.}~\bibnamefont {Beth~Rothwell}}, \bibinfo {author} {\bibfnamefont {G.~A.}\ \bibnamefont {Keefe}}, \bibinfo {author} {\bibfnamefont {M.}~\bibnamefont {Steffen}},\ and\ \bibinfo {author} {\bibfnamefont {D.~P.}\ \bibnamefont {Pappas}},\ }\bibfield  {title} {\enquote {\bibinfo {title} {Improved superconducting qubit coherence using titanium nitride},}\ }\href {https://doi.org/10.1063/1.4813269} {\bibfield  {journal} {\bibinfo  {journal}
  {Applied Physics Letters}\ }\textbf {\bibinfo {volume} {103}},\ \bibinfo {pages} {012602} (\bibinfo {year} {2013})}\BibitemShut {NoStop}%
\bibitem [{\citenamefont {Deng}\ \emph {et~al.}(2023)\citenamefont {Deng}, \citenamefont {Song}, \citenamefont {Gao}, \citenamefont {Xia}, \citenamefont {Bao}, \citenamefont {Jiang}, \citenamefont {Ku}, \citenamefont {Li}, \citenamefont {Ma}, \citenamefont {Qin}, \citenamefont {Sun}, \citenamefont {Tang}, \citenamefont {Wang}, \citenamefont {Wu}, \citenamefont {Yu}, \citenamefont {Zhang}, \citenamefont {Zhang}, \citenamefont {Zhou}, \citenamefont {Zhu}, \citenamefont {Shi}, \citenamefont {Zhao},\ and\ \citenamefont {Deng}}]{Deng2023Feb_TiN}%
  \BibitemOpen
  \bibfield  {author} {\bibinfo {author} {\bibfnamefont {H.}~\bibnamefont {Deng}}, \bibinfo {author} {\bibfnamefont {Z.}~\bibnamefont {Song}}, \bibinfo {author} {\bibfnamefont {R.}~\bibnamefont {Gao}}, \bibinfo {author} {\bibfnamefont {T.}~\bibnamefont {Xia}}, \bibinfo {author} {\bibfnamefont {F.}~\bibnamefont {Bao}}, \bibinfo {author} {\bibfnamefont {X.}~\bibnamefont {Jiang}}, \bibinfo {author} {\bibfnamefont {H.-S.}\ \bibnamefont {Ku}}, \bibinfo {author} {\bibfnamefont {Z.}~\bibnamefont {Li}}, \bibinfo {author} {\bibfnamefont {X.}~\bibnamefont {Ma}}, \bibinfo {author} {\bibfnamefont {J.}~\bibnamefont {Qin}}, \bibinfo {author} {\bibfnamefont {H.}~\bibnamefont {Sun}}, \bibinfo {author} {\bibfnamefont {C.}~\bibnamefont {Tang}}, \bibinfo {author} {\bibfnamefont {T.}~\bibnamefont {Wang}}, \bibinfo {author} {\bibfnamefont {F.}~\bibnamefont {Wu}}, \bibinfo {author} {\bibfnamefont {W.}~\bibnamefont {Yu}}, \bibinfo {author} {\bibfnamefont {G.}~\bibnamefont {Zhang}}, \bibinfo {author} {\bibfnamefont {X.}~\bibnamefont
  {Zhang}}, \bibinfo {author} {\bibfnamefont {J.}~\bibnamefont {Zhou}}, \bibinfo {author} {\bibfnamefont {X.}~\bibnamefont {Zhu}}, \bibinfo {author} {\bibfnamefont {Y.}~\bibnamefont {Shi}}, \bibinfo {author} {\bibfnamefont {H.-H.}\ \bibnamefont {Zhao}},\ and\ \bibinfo {author} {\bibfnamefont {C.}~\bibnamefont {Deng}},\ }\bibfield  {title} {\enquote {\bibinfo {title} {Titanium nitride film on sapphire substrate with low dielectric loss for superconducting qubits},}\ }\href {https://doi.org/10.1103/PhysRevApplied.19.024013} {\bibfield  {journal} {\bibinfo  {journal} {Phys. Rev. Appl.}\ }\textbf {\bibinfo {volume} {19}},\ \bibinfo {pages} {024013} (\bibinfo {year} {2023})}\BibitemShut {NoStop}%
\bibitem [{\citenamefont {Read}\ and\ \citenamefont {Altman}(1965)}]{Read1965Aug}%
  \BibitemOpen
  \bibfield  {author} {\bibinfo {author} {\bibfnamefont {M.~H.}\ \bibnamefont {Read}}\ and\ \bibinfo {author} {\bibfnamefont {C.}~\bibnamefont {Altman}},\ }\bibfield  {title} {\enquote {\bibinfo {title} {{A new structure in Tantalum thin films}},}\ }\href {https://doi.org/10.1063/1.1754294} {\bibfield  {journal} {\bibinfo  {journal} {Appl. Phys. Lett.}\ }\textbf {\bibinfo {volume} {7}},\ \bibinfo {pages} {51--52} (\bibinfo {year} {1965})}\BibitemShut {NoStop}%
\bibitem [{\citenamefont {Schwartz}\ \emph {et~al.}(1972)\citenamefont {Schwartz}, \citenamefont {Reed}, \citenamefont {Polash},\ and\ \citenamefont {Read}}]{SCHWARTZ1972333}%
  \BibitemOpen
  \bibfield  {author} {\bibinfo {author} {\bibfnamefont {N.}~\bibnamefont {Schwartz}}, \bibinfo {author} {\bibfnamefont {W.}~\bibnamefont {Reed}}, \bibinfo {author} {\bibfnamefont {P.}~\bibnamefont {Polash}},\ and\ \bibinfo {author} {\bibfnamefont {M.~H.}\ \bibnamefont {Read}},\ }\bibfield  {title} {\enquote {\bibinfo {title} {Temperature coefficient of resistance of beta-tantalum films and mixtures with b.c.c.-tantalum},}\ }\href {https://doi.org/https://doi.org/10.1016/0040-6090(72)90433-6} {\bibfield  {journal} {\bibinfo  {journal} {Thin Solid Films}\ }\textbf {\bibinfo {volume} {14}},\ \bibinfo {pages} {333--346} (\bibinfo {year} {1972})}\BibitemShut {NoStop}%
\bibitem [{\citenamefont {Navid}\ and\ \citenamefont {Hodge}(2012)}]{Navid2012Feb}%
  \BibitemOpen
  \bibfield  {author} {\bibinfo {author} {\bibfnamefont {A.~A.}\ \bibnamefont {Navid}}\ and\ \bibinfo {author} {\bibfnamefont {A.~M.}\ \bibnamefont {Hodge}},\ }\bibfield  {title} {\enquote {\bibinfo {title} {{Nanostructured alpha and beta tantalum formation{\ifmmode---\else\textemdash\fi}Relationship between plasma parameters and microstructure}},}\ }\href {https://doi.org/10.1016/j.msea.2011.12.017} {\bibfield  {journal} {\bibinfo  {journal} {Mater. Sci. Eng., A}\ }\textbf {\bibinfo {volume} {536}},\ \bibinfo {pages} {49--56} (\bibinfo {year} {2012})}\BibitemShut {NoStop}%
\bibitem [{\citenamefont {Ganjam}\ \emph {et~al.}(5 01)\citenamefont {Ganjam}, \citenamefont {Wang}, \citenamefont {Lu}, \citenamefont {Banerjee}, \citenamefont {Lei}, \citenamefont {Krayzman}, \citenamefont {Kisslinger}, \citenamefont {Zhou}, \citenamefont {Li}, \citenamefont {Jia}, \citenamefont {Liu}, \citenamefont {Frunzio},\ and\ \citenamefont {Schoelkopf}}]{ganjam_surpassing_2024}%
  \BibitemOpen
  \bibfield  {author} {\bibinfo {author} {\bibfnamefont {S.}~\bibnamefont {Ganjam}}, \bibinfo {author} {\bibfnamefont {Y.}~\bibnamefont {Wang}}, \bibinfo {author} {\bibfnamefont {Y.}~\bibnamefont {Lu}}, \bibinfo {author} {\bibfnamefont {A.}~\bibnamefont {Banerjee}}, \bibinfo {author} {\bibfnamefont {C.~U.}\ \bibnamefont {Lei}}, \bibinfo {author} {\bibfnamefont {L.}~\bibnamefont {Krayzman}}, \bibinfo {author} {\bibfnamefont {K.}~\bibnamefont {Kisslinger}}, \bibinfo {author} {\bibfnamefont {C.}~\bibnamefont {Zhou}}, \bibinfo {author} {\bibfnamefont {R.}~\bibnamefont {Li}}, \bibinfo {author} {\bibfnamefont {Y.}~\bibnamefont {Jia}}, \bibinfo {author} {\bibfnamefont {M.}~\bibnamefont {Liu}}, \bibinfo {author} {\bibfnamefont {L.}~\bibnamefont {Frunzio}},\ and\ \bibinfo {author} {\bibfnamefont {R.~J.}\ \bibnamefont {Schoelkopf}},\ }\bibfield  {title} {\enquote {\bibinfo {title} {Surpassing millisecond coherence in on chip superconducting quantum memories by optimizing materials and circuit design.}}\ }\href
  {https://doi.org/10.1038/s41467-024-47857-6} {\bibfield  {journal} {\bibinfo  {journal} {Nat Commun}\ }\textbf {\bibinfo {volume} {15}},\ \bibinfo {pages} {3687} (\bibinfo {year} {2024-05-01})}\BibitemShut {NoStop}%
\bibitem [{\citenamefont {McLellan}\ \emph {et~al.}(2023)\citenamefont {McLellan}, \citenamefont {Dutta}, \citenamefont {Zhou}, \citenamefont {Jia}, \citenamefont {Weiland}, \citenamefont {Gui}, \citenamefont {Place}, \citenamefont {Crowley}, \citenamefont {Le}, \citenamefont {Madhavan}, \citenamefont {Gang}, \citenamefont {Baker}, \citenamefont {Head}, \citenamefont {Waluyo}, \citenamefont {Li}, \citenamefont {Kisslinger}, \citenamefont {Hunt}, \citenamefont {Jarrige}, \citenamefont {Lyon}, \citenamefont {Barbour}, \citenamefont {Cava}, \citenamefont {Houck}, \citenamefont {Hulbert}, \citenamefont {Liu}, \citenamefont {Walter},\ and\ \citenamefont {de~Leon}}]{McLellan2023Jul}%
  \BibitemOpen
  \bibfield  {author} {\bibinfo {author} {\bibfnamefont {R.~A.}\ \bibnamefont {McLellan}}, \bibinfo {author} {\bibfnamefont {A.}~\bibnamefont {Dutta}}, \bibinfo {author} {\bibfnamefont {C.}~\bibnamefont {Zhou}}, \bibinfo {author} {\bibfnamefont {Y.}~\bibnamefont {Jia}}, \bibinfo {author} {\bibfnamefont {C.}~\bibnamefont {Weiland}}, \bibinfo {author} {\bibfnamefont {X.}~\bibnamefont {Gui}}, \bibinfo {author} {\bibfnamefont {A.~P.~M.}\ \bibnamefont {Place}}, \bibinfo {author} {\bibfnamefont {K.~D.}\ \bibnamefont {Crowley}}, \bibinfo {author} {\bibfnamefont {X.~H.}\ \bibnamefont {Le}}, \bibinfo {author} {\bibfnamefont {T.}~\bibnamefont {Madhavan}}, \bibinfo {author} {\bibfnamefont {Y.}~\bibnamefont {Gang}}, \bibinfo {author} {\bibfnamefont {L.}~\bibnamefont {Baker}}, \bibinfo {author} {\bibfnamefont {A.~R.}\ \bibnamefont {Head}}, \bibinfo {author} {\bibfnamefont {I.}~\bibnamefont {Waluyo}}, \bibinfo {author} {\bibfnamefont {R.}~\bibnamefont {Li}}, \bibinfo {author} {\bibfnamefont {K.}~\bibnamefont {Kisslinger}},
  \bibinfo {author} {\bibfnamefont {A.}~\bibnamefont {Hunt}}, \bibinfo {author} {\bibfnamefont {I.}~\bibnamefont {Jarrige}}, \bibinfo {author} {\bibfnamefont {S.~A.}\ \bibnamefont {Lyon}}, \bibinfo {author} {\bibfnamefont {A.~M.}\ \bibnamefont {Barbour}}, \bibinfo {author} {\bibfnamefont {R.~J.}\ \bibnamefont {Cava}}, \bibinfo {author} {\bibfnamefont {A.~A.}\ \bibnamefont {Houck}}, \bibinfo {author} {\bibfnamefont {S.~L.}\ \bibnamefont {Hulbert}}, \bibinfo {author} {\bibfnamefont {M.}~\bibnamefont {Liu}}, \bibinfo {author} {\bibfnamefont {A.~L.}\ \bibnamefont {Walter}},\ and\ \bibinfo {author} {\bibfnamefont {N.~P.}\ \bibnamefont {de~Leon}},\ }\bibfield  {title} {\enquote {\bibinfo {title} {{Chemical Profiles of the Oxides on Tantalum in State of the Art Superconducting Circuits}},}\ }\href {https://doi.org/10.1002/advs.202300921} {\bibfield  {journal} {\bibinfo  {journal} {Adv. Sci.}\ }\textbf {\bibinfo {volume} {10}},\ \bibinfo {pages} {2300921} (\bibinfo {year} {2023})}\BibitemShut {NoStop}%
\bibitem [{\citenamefont {Bal}\ \emph {et~al.}(2024)\citenamefont {Bal}, \citenamefont {Murthy}, \citenamefont {Zhu}, \citenamefont {Crisa}, \citenamefont {You}, \citenamefont {Huang}, \citenamefont {Roy}, \citenamefont {Lee}, \citenamefont {Zanten}, \citenamefont {Pilipenko}, \citenamefont {Nekrashevich}, \citenamefont {Lunin}, \citenamefont {Bafia}, \citenamefont {Krasnikova}, \citenamefont {Kopas}, \citenamefont {Lachman}, \citenamefont {Miller}, \citenamefont {Mutus}, \citenamefont {Reagor}, \citenamefont {Cansizoglu}, \citenamefont {Marshall}, \citenamefont {Pappas}, \citenamefont {Vu}, \citenamefont {Yadavalli}, \citenamefont {Oh}, \citenamefont {Zhou}, \citenamefont {Kramer}, \citenamefont {Lecocq}, \citenamefont {Goronzy}, \citenamefont {Torres-Castanedo}, \citenamefont {Pritchard}, \citenamefont {Dravid}, \citenamefont {Rondinelli}, \citenamefont {Bedzyk}, \citenamefont {Hersam}, \citenamefont {Zasadzinski}, \citenamefont {Koch}, \citenamefont {Sauls}, \citenamefont {Romanenko},\ and\ \citenamefont
  {Grassellino}}]{Mustafa2024FermiLab}%
  \BibitemOpen
  \bibfield  {author} {\bibinfo {author} {\bibfnamefont {M.}~\bibnamefont {Bal}}, \bibinfo {author} {\bibfnamefont {A.~A.}\ \bibnamefont {Murthy}}, \bibinfo {author} {\bibfnamefont {S.}~\bibnamefont {Zhu}}, \bibinfo {author} {\bibfnamefont {F.}~\bibnamefont {Crisa}}, \bibinfo {author} {\bibfnamefont {X.}~\bibnamefont {You}}, \bibinfo {author} {\bibfnamefont {Z.}~\bibnamefont {Huang}}, \bibinfo {author} {\bibfnamefont {T.}~\bibnamefont {Roy}}, \bibinfo {author} {\bibfnamefont {J.}~\bibnamefont {Lee}}, \bibinfo {author} {\bibfnamefont {D.~v.}\ \bibnamefont {Zanten}}, \bibinfo {author} {\bibfnamefont {R.}~\bibnamefont {Pilipenko}}, \bibinfo {author} {\bibfnamefont {I.}~\bibnamefont {Nekrashevich}}, \bibinfo {author} {\bibfnamefont {A.}~\bibnamefont {Lunin}}, \bibinfo {author} {\bibfnamefont {D.}~\bibnamefont {Bafia}}, \bibinfo {author} {\bibfnamefont {Y.}~\bibnamefont {Krasnikova}}, \bibinfo {author} {\bibfnamefont {C.~J.}\ \bibnamefont {Kopas}}, \bibinfo {author} {\bibfnamefont {E.~O.}\ \bibnamefont {Lachman}},
  \bibinfo {author} {\bibfnamefont {D.}~\bibnamefont {Miller}}, \bibinfo {author} {\bibfnamefont {J.~Y.}\ \bibnamefont {Mutus}}, \bibinfo {author} {\bibfnamefont {M.~J.}\ \bibnamefont {Reagor}}, \bibinfo {author} {\bibfnamefont {H.}~\bibnamefont {Cansizoglu}}, \bibinfo {author} {\bibfnamefont {J.}~\bibnamefont {Marshall}}, \bibinfo {author} {\bibfnamefont {D.~P.}\ \bibnamefont {Pappas}}, \bibinfo {author} {\bibfnamefont {K.}~\bibnamefont {Vu}}, \bibinfo {author} {\bibfnamefont {K.}~\bibnamefont {Yadavalli}}, \bibinfo {author} {\bibfnamefont {J.-S.}\ \bibnamefont {Oh}}, \bibinfo {author} {\bibfnamefont {L.}~\bibnamefont {Zhou}}, \bibinfo {author} {\bibfnamefont {M.~J.}\ \bibnamefont {Kramer}}, \bibinfo {author} {\bibfnamefont {F.}~\bibnamefont {Lecocq}}, \bibinfo {author} {\bibfnamefont {D.~P.}\ \bibnamefont {Goronzy}}, \bibinfo {author} {\bibfnamefont {C.~G.}\ \bibnamefont {Torres-Castanedo}}, \bibinfo {author} {\bibfnamefont {P.~G.}\ \bibnamefont {Pritchard}}, \bibinfo {author} {\bibfnamefont {V.~P.}\
  \bibnamefont {Dravid}}, \bibinfo {author} {\bibfnamefont {J.~M.}\ \bibnamefont {Rondinelli}}, \bibinfo {author} {\bibfnamefont {M.~J.}\ \bibnamefont {Bedzyk}}, \bibinfo {author} {\bibfnamefont {M.~C.}\ \bibnamefont {Hersam}}, \bibinfo {author} {\bibfnamefont {J.}~\bibnamefont {Zasadzinski}}, \bibinfo {author} {\bibfnamefont {J.}~\bibnamefont {Koch}}, \bibinfo {author} {\bibfnamefont {J.~A.}\ \bibnamefont {Sauls}}, \bibinfo {author} {\bibfnamefont {A.}~\bibnamefont {Romanenko}},\ and\ \bibinfo {author} {\bibfnamefont {A.}~\bibnamefont {Grassellino}},\ }\bibfield  {title} {\enquote {\bibinfo {title} {Systematic improvements in transmon qubit coherence enabled by niobium surface encapsulation},}\ }\href {https://doi.org/10.1038/s41534-024-00840-x} {\bibfield  {journal} {\bibinfo  {journal} {npj Quantum Information}\ }\textbf {\bibinfo {volume} {10}},\ \bibinfo {pages} {43} (\bibinfo {year} {2024})}\BibitemShut {NoStop}%
\bibitem [{\citenamefont {Lozano}\ \emph {et~al.}(2024)\citenamefont {Lozano}, \citenamefont {Mongillo}, \citenamefont {Piao}, \citenamefont {Couet}, \citenamefont {Wan}, \citenamefont {Canvel}, \citenamefont {Vadiraj}, \citenamefont {Ivanov}, \citenamefont {Verjauw}, \citenamefont {Acharya}, \citenamefont {Damme}, \citenamefont {Mohiyaddin}, \citenamefont {Jussot}, \citenamefont {Gowda}, \citenamefont {Pacco}, \citenamefont {Raes}, \citenamefont {de~Vondel}, \citenamefont {Radu}, \citenamefont {Govoreanu}, \citenamefont {Swerts}, \citenamefont {Potočnik},\ and\ \citenamefont {Greve}}]{Lozano2024Ta_on_silicon}%
  \BibitemOpen
  \bibfield  {author} {\bibinfo {author} {\bibfnamefont {D.~P.}\ \bibnamefont {Lozano}}, \bibinfo {author} {\bibfnamefont {M.}~\bibnamefont {Mongillo}}, \bibinfo {author} {\bibfnamefont {X.}~\bibnamefont {Piao}}, \bibinfo {author} {\bibfnamefont {S.}~\bibnamefont {Couet}}, \bibinfo {author} {\bibfnamefont {D.}~\bibnamefont {Wan}}, \bibinfo {author} {\bibfnamefont {Y.}~\bibnamefont {Canvel}}, \bibinfo {author} {\bibfnamefont {A.~M.}\ \bibnamefont {Vadiraj}}, \bibinfo {author} {\bibfnamefont {T.}~\bibnamefont {Ivanov}}, \bibinfo {author} {\bibfnamefont {J.}~\bibnamefont {Verjauw}}, \bibinfo {author} {\bibfnamefont {R.}~\bibnamefont {Acharya}}, \bibinfo {author} {\bibfnamefont {J.~V.}\ \bibnamefont {Damme}}, \bibinfo {author} {\bibfnamefont {F.~A.}\ \bibnamefont {Mohiyaddin}}, \bibinfo {author} {\bibfnamefont {J.}~\bibnamefont {Jussot}}, \bibinfo {author} {\bibfnamefont {P.~P.}\ \bibnamefont {Gowda}}, \bibinfo {author} {\bibfnamefont {A.}~\bibnamefont {Pacco}}, \bibinfo {author} {\bibfnamefont {B.}~\bibnamefont
  {Raes}}, \bibinfo {author} {\bibfnamefont {J.~V.}\ \bibnamefont {de~Vondel}}, \bibinfo {author} {\bibfnamefont {I.~P.}\ \bibnamefont {Radu}}, \bibinfo {author} {\bibfnamefont {B.}~\bibnamefont {Govoreanu}}, \bibinfo {author} {\bibfnamefont {J.}~\bibnamefont {Swerts}}, \bibinfo {author} {\bibfnamefont {A.}~\bibnamefont {Potočnik}},\ and\ \bibinfo {author} {\bibfnamefont {K.~D.}\ \bibnamefont {Greve}},\ }\bibfield  {title} {\enquote {\bibinfo {title} {Low-loss $\alpha$-tantalum coplanar waveguide resonators on silicon wafers: fabrication, characterization and surface modification},}\ }\href {https://doi.org/10.1088/2633-4356/ad4b8c} {\bibfield  {journal} {\bibinfo  {journal} {Materials for Quantum Technology}\ }\textbf {\bibinfo {volume} {4}},\ \bibinfo {pages} {025801} (\bibinfo {year} {2024})}\BibitemShut {NoStop}%
\bibitem [{\citenamefont {Kobler}\ \emph {et~al.}(2013)\citenamefont {Kobler}, \citenamefont {Kashiwar}, \citenamefont {Hahn},\ and\ \citenamefont {Kübel}}]{Kobler2013ACOM}%
  \BibitemOpen
  \bibfield  {author} {\bibinfo {author} {\bibfnamefont {A.}~\bibnamefont {Kobler}}, \bibinfo {author} {\bibfnamefont {A.}~\bibnamefont {Kashiwar}}, \bibinfo {author} {\bibfnamefont {H.}~\bibnamefont {Hahn}},\ and\ \bibinfo {author} {\bibfnamefont {C.}~\bibnamefont {Kübel}},\ }\bibfield  {title} {\enquote {\bibinfo {title} {Combination of in situ straining and acom tem: A novel method for analysis of plastic deformation of nanocrystalline metals},}\ }\href {https://doi.org/https://doi.org/10.1016/j.ultramic.2012.12.019} {\bibfield  {journal} {\bibinfo  {journal} {Ultramicroscopy}\ }\textbf {\bibinfo {volume} {128}},\ \bibinfo {pages} {68--81} (\bibinfo {year} {2013})}\BibitemShut {NoStop}%
\bibitem [{\citenamefont {Hauser}\ and\ \citenamefont {Theuerer}(1964)}]{Hauser1964Apr}%
  \BibitemOpen
  \bibfield  {author} {\bibinfo {author} {\bibfnamefont {J.~J.}\ \bibnamefont {Hauser}}\ and\ \bibinfo {author} {\bibfnamefont {H.~C.}\ \bibnamefont {Theuerer}},\ }\bibfield  {title} {\enquote {\bibinfo {title} {{Size Effects in Thin Films of ${\mathrm{V}}_{3}$Ge, Nb, and Ta}},}\ }\href {https://doi.org/10.1103/PhysRev.134.A198} {\bibfield  {journal} {\bibinfo  {journal} {Phys. Rev.}\ }\textbf {\bibinfo {volume} {134}},\ \bibinfo {pages} {A198--A205} (\bibinfo {year} {1964})}\BibitemShut {NoStop}%
\bibitem [{\citenamefont {Balashov}, \citenamefont {Meyer},\ and\ \citenamefont {Wulfhekel}(2018)}]{balashov_compact_2018}%
  \BibitemOpen
  \bibfield  {author} {\bibinfo {author} {\bibfnamefont {T.}~\bibnamefont {Balashov}}, \bibinfo {author} {\bibfnamefont {M.}~\bibnamefont {Meyer}},\ and\ \bibinfo {author} {\bibfnamefont {W.}~\bibnamefont {Wulfhekel}},\ }\bibfield  {title} {\enquote {\bibinfo {title} {A compact ultrahigh vacuum scanning tunneling microscope with dilution refrigeration},}\ }\href {https://doi.org/10.1063/1.5043636} {\bibfield  {journal} {\bibinfo  {journal} {Review of Scientific Instruments}\ }\textbf {\bibinfo {volume} {89}},\ \bibinfo {pages} {113707} (\bibinfo {year} {2018})}\BibitemShut {NoStop}%
\bibitem [{\citenamefont {Reed}\ and\ \citenamefont {Boyer}(1976)}]{Reed1976Jul}%
  \BibitemOpen
  \bibfield  {author} {\bibinfo {author} {\bibfnamefont {R.~W.}\ \bibnamefont {Reed}}\ and\ \bibinfo {author} {\bibfnamefont {A.~C.}\ \bibnamefont {Boyer}},\ }\bibfield  {title} {\enquote {\bibinfo {title} {{Ultrasonic determination of the superconducting energy gap in high-purity tantalum}},}\ }\href {https://doi.org/10.1007/BF00659193} {\bibfield  {journal} {\bibinfo  {journal} {J. Low Temp. Phys.}\ }\textbf {\bibinfo {volume} {24}},\ \bibinfo {pages} {35--40} (\bibinfo {year} {1976})}\BibitemShut {NoStop}%
\bibitem [{\citenamefont {Bardeen}, \citenamefont {Cooper},\ and\ \citenamefont {Schrieffer}(1957)}]{Bardeen1957}%
  \BibitemOpen
  \bibfield  {author} {\bibinfo {author} {\bibfnamefont {J.}~\bibnamefont {Bardeen}}, \bibinfo {author} {\bibfnamefont {L.~N.}\ \bibnamefont {Cooper}},\ and\ \bibinfo {author} {\bibfnamefont {J.~R.}\ \bibnamefont {Schrieffer}},\ }\bibfield  {title} {\enquote {\bibinfo {title} {Theory of superconductivity},}\ }\href {https://doi.org/10.1103/PhysRev.108.1175} {\bibfield  {journal} {\bibinfo  {journal} {Phys. Rev.}\ }\textbf {\bibinfo {volume} {108}},\ \bibinfo {pages} {1175--1204} (\bibinfo {year} {1957})}\BibitemShut {NoStop}%
\bibitem [{\citenamefont {Arabi}\ \emph {et~al.}(2024)\citenamefont {Arabi}, \citenamefont {Li}, \citenamefont {Dhundhwal}, \citenamefont {Fuchs}, \citenamefont {Reisinger}, \citenamefont {Pop},\ and\ \citenamefont {Wulfhekel}}]{Arabi2024Dec}%
  \BibitemOpen
  \bibfield  {author} {\bibinfo {author} {\bibfnamefont {S.}~\bibnamefont {Arabi}}, \bibinfo {author} {\bibfnamefont {Q.}~\bibnamefont {Li}}, \bibinfo {author} {\bibfnamefont {R.}~\bibnamefont {Dhundhwal}}, \bibinfo {author} {\bibfnamefont {D.}~\bibnamefont {Fuchs}}, \bibinfo {author} {\bibfnamefont {T.}~\bibnamefont {Reisinger}}, \bibinfo {author} {\bibfnamefont {I.~M.}\ \bibnamefont {Pop}},\ and\ \bibinfo {author} {\bibfnamefont {W.}~\bibnamefont {Wulfhekel}},\ }\bibfield  {title} {\enquote {\bibinfo {title} {{Magnetic bound states embedded in tantalum superconducting thin films}},}\ }\href {https://doi.org/10.48550/arXiv.2412.15903} {\bibfield  {journal} {\bibinfo  {journal} {arXiv}\ } (\bibinfo {year} {2024}),\ 10.48550/arXiv.2412.15903},\ \Eprint {https://arxiv.org/abs/2412.15903} {2412.15903} \BibitemShut {NoStop}%
\bibitem [{\citenamefont {Moseley}\ and\ \citenamefont {Seabrook}(1973)}]{Moseley1973}%
  \BibitemOpen
  \bibfield  {author} {\bibinfo {author} {\bibfnamefont {P.~T.}\ \bibnamefont {Moseley}}\ and\ \bibinfo {author} {\bibfnamefont {C.~J.}\ \bibnamefont {Seabrook}},\ }\bibfield  {title} {\enquote {\bibinfo {title} {{The crystal structure of {$\beta$}-tantalum}},}\ }\href {https://doi.org/10.1107/S0567740873004140} {\bibfield  {journal} {\bibinfo  {journal} {Acta Crystallographica Section B}\ }\textbf {\bibinfo {volume} {29}},\ \bibinfo {pages} {1170--1171} (\bibinfo {year} {1973})}\BibitemShut {NoStop}%
\bibitem [{\citenamefont {Jain}\ \emph {et~al.}(2013)\citenamefont {Jain}, \citenamefont {Ong}, \citenamefont {Hautier}, \citenamefont {Chen}, \citenamefont {Richards}, \citenamefont {Dacek}, \citenamefont {Cholia}, \citenamefont {Gunter}, \citenamefont {Skinner}, \citenamefont {Ceder},\ and\ \citenamefont {Persson}}]{materialsproject2013}%
  \BibitemOpen
  \bibfield  {author} {\bibinfo {author} {\bibfnamefont {A.}~\bibnamefont {Jain}}, \bibinfo {author} {\bibfnamefont {S.~P.}\ \bibnamefont {Ong}}, \bibinfo {author} {\bibfnamefont {G.}~\bibnamefont {Hautier}}, \bibinfo {author} {\bibfnamefont {W.}~\bibnamefont {Chen}}, \bibinfo {author} {\bibfnamefont {W.~D.}\ \bibnamefont {Richards}}, \bibinfo {author} {\bibfnamefont {S.}~\bibnamefont {Dacek}}, \bibinfo {author} {\bibfnamefont {S.}~\bibnamefont {Cholia}}, \bibinfo {author} {\bibfnamefont {D.}~\bibnamefont {Gunter}}, \bibinfo {author} {\bibfnamefont {D.}~\bibnamefont {Skinner}}, \bibinfo {author} {\bibfnamefont {G.}~\bibnamefont {Ceder}},\ and\ \bibinfo {author} {\bibfnamefont {K.~A.}\ \bibnamefont {Persson}},\ }\bibfield  {title} {\enquote {\bibinfo {title} {{Commentary: The Materials Project: A materials genome approach to accelerating materials innovation}},}\ }\href {https://doi.org/10.1063/1.4812323} {\bibfield  {journal} {\bibinfo  {journal} {APL Mater.}\ }\textbf {\bibinfo {volume} {1}} (\bibinfo {year}
  {2013}),\ 10.1063/1.4812323}\BibitemShut {NoStop}%
\bibitem [{\citenamefont {Arakcheeva}\ \emph {et~al.}(2003)\citenamefont {Arakcheeva}, \citenamefont {Chapuis}, \citenamefont {Birkedal}, \citenamefont {Pattison},\ and\ \citenamefont {Grinevitch}}]{Arakcheeva:sn0032}%
  \BibitemOpen
  \bibfield  {author} {\bibinfo {author} {\bibfnamefont {A.}~\bibnamefont {Arakcheeva}}, \bibinfo {author} {\bibfnamefont {G.}~\bibnamefont {Chapuis}}, \bibinfo {author} {\bibfnamefont {H.}~\bibnamefont {Birkedal}}, \bibinfo {author} {\bibfnamefont {P.}~\bibnamefont {Pattison}},\ and\ \bibinfo {author} {\bibfnamefont {V.}~\bibnamefont {Grinevitch}},\ }\bibfield  {title} {\enquote {\bibinfo {title} {{The commensurate composite {$\sigma$}-structure of {$\beta$}-tantalum}},}\ }\href {https://doi.org/10.1107/S0108768103009005} {\bibfield  {journal} {\bibinfo  {journal} {Acta Crystallographica Section B}\ }\textbf {\bibinfo {volume} {59}},\ \bibinfo {pages} {324--336} (\bibinfo {year} {2003})}\BibitemShut {NoStop}%
\bibitem [{\citenamefont {Rieger}\ \emph {et~al.}(2023)\citenamefont {Rieger}, \citenamefont {G\"unzler}, \citenamefont {Spiecker}, \citenamefont {Nambisan}, \citenamefont {Wernsdorfer},\ and\ \citenamefont {Pop}}]{Rieger2023Fano}%
  \BibitemOpen
  \bibfield  {author} {\bibinfo {author} {\bibfnamefont {D.}~\bibnamefont {Rieger}}, \bibinfo {author} {\bibfnamefont {S.}~\bibnamefont {G\"unzler}}, \bibinfo {author} {\bibfnamefont {M.}~\bibnamefont {Spiecker}}, \bibinfo {author} {\bibfnamefont {A.}~\bibnamefont {Nambisan}}, \bibinfo {author} {\bibfnamefont {W.}~\bibnamefont {Wernsdorfer}},\ and\ \bibinfo {author} {\bibfnamefont {I.}~\bibnamefont {Pop}},\ }\bibfield  {title} {\enquote {\bibinfo {title} {Fano interference in microwave resonator measurements},}\ }\href {https://doi.org/10.1103/PhysRevApplied.20.014059} {\bibfield  {journal} {\bibinfo  {journal} {Phys. Rev. Appl.}\ }\textbf {\bibinfo {volume} {20}},\ \bibinfo {pages} {014059} (\bibinfo {year} {2023})}\BibitemShut {NoStop}%
\bibitem [{\citenamefont {Winkel}\ \emph {et~al.}(2020)\citenamefont {Winkel}, \citenamefont {Borisov}, \citenamefont {Gr\"unhaupt}, \citenamefont {Rieger}, \citenamefont {Spiecker}, \citenamefont {Valenti}, \citenamefont {Ustinov}, \citenamefont {Wernsdorfer},\ and\ \citenamefont {Pop}}]{Winkel2020}%
  \BibitemOpen
  \bibfield  {author} {\bibinfo {author} {\bibfnamefont {P.}~\bibnamefont {Winkel}}, \bibinfo {author} {\bibfnamefont {K.}~\bibnamefont {Borisov}}, \bibinfo {author} {\bibfnamefont {L.}~\bibnamefont {Gr\"unhaupt}}, \bibinfo {author} {\bibfnamefont {D.}~\bibnamefont {Rieger}}, \bibinfo {author} {\bibfnamefont {M.}~\bibnamefont {Spiecker}}, \bibinfo {author} {\bibfnamefont {F.}~\bibnamefont {Valenti}}, \bibinfo {author} {\bibfnamefont {A.~V.}\ \bibnamefont {Ustinov}}, \bibinfo {author} {\bibfnamefont {W.}~\bibnamefont {Wernsdorfer}},\ and\ \bibinfo {author} {\bibfnamefont {I.~M.}\ \bibnamefont {Pop}},\ }\bibfield  {title} {\enquote {\bibinfo {title} {Implementation of a transmon qubit using superconducting granular aluminum},}\ }\href {https://doi.org/10.1103/PhysRevX.10.031032} {\bibfield  {journal} {\bibinfo  {journal} {Phys. Rev. X}\ }\textbf {\bibinfo {volume} {10}},\ \bibinfo {pages} {031032} (\bibinfo {year} {2020})}\BibitemShut {NoStop}%
\bibitem [{\citenamefont {Gr\"unhaupt}\ \emph {et~al.}(2018)\citenamefont {Gr\"unhaupt}, \citenamefont {Maleeva}, \citenamefont {Skacel}, \citenamefont {Calvo}, \citenamefont {Levy-Bertrand}, \citenamefont {Ustinov}, \citenamefont {Rotzinger}, \citenamefont {Monfardini}, \citenamefont {Catelani},\ and\ \citenamefont {Pop}}]{LukasG2018}%
  \BibitemOpen
  \bibfield  {author} {\bibinfo {author} {\bibfnamefont {L.}~\bibnamefont {Gr\"unhaupt}}, \bibinfo {author} {\bibfnamefont {N.}~\bibnamefont {Maleeva}}, \bibinfo {author} {\bibfnamefont {S.~T.}\ \bibnamefont {Skacel}}, \bibinfo {author} {\bibfnamefont {M.}~\bibnamefont {Calvo}}, \bibinfo {author} {\bibfnamefont {F.}~\bibnamefont {Levy-Bertrand}}, \bibinfo {author} {\bibfnamefont {A.~V.}\ \bibnamefont {Ustinov}}, \bibinfo {author} {\bibfnamefont {H.}~\bibnamefont {Rotzinger}}, \bibinfo {author} {\bibfnamefont {A.}~\bibnamefont {Monfardini}}, \bibinfo {author} {\bibfnamefont {G.}~\bibnamefont {Catelani}},\ and\ \bibinfo {author} {\bibfnamefont {I.~M.}\ \bibnamefont {Pop}},\ }\bibfield  {title} {\enquote {\bibinfo {title} {Loss mechanisms and quasiparticle dynamics in superconducting microwave resonators made of thin-film granular aluminum},}\ }\href {https://doi.org/10.1103/PhysRevLett.121.117001} {\bibfield  {journal} {\bibinfo  {journal} {Phys. Rev. Lett.}\ }\textbf {\bibinfo {volume} {121}},\ \bibinfo
  {pages} {117001} (\bibinfo {year} {2018})}\BibitemShut {NoStop}%
\bibitem [{\citenamefont {de~Visser}\ \emph {et~al.}(2014)\citenamefont {de~Visser}, \citenamefont {Goldie}, \citenamefont {Diener}, \citenamefont {Withington}, \citenamefont {Baselmans},\ and\ \citenamefont {Klapwijk}}]{Visser2014}%
  \BibitemOpen
  \bibfield  {author} {\bibinfo {author} {\bibfnamefont {P.~J.}\ \bibnamefont {de~Visser}}, \bibinfo {author} {\bibfnamefont {D.~J.}\ \bibnamefont {Goldie}}, \bibinfo {author} {\bibfnamefont {P.}~\bibnamefont {Diener}}, \bibinfo {author} {\bibfnamefont {S.}~\bibnamefont {Withington}}, \bibinfo {author} {\bibfnamefont {J.~J.~A.}\ \bibnamefont {Baselmans}},\ and\ \bibinfo {author} {\bibfnamefont {T.~M.}\ \bibnamefont {Klapwijk}},\ }\bibfield  {title} {\enquote {\bibinfo {title} {Evidence of a nonequilibrium distribution of quasiparticles in the microwave response of a superconducting aluminum resonator},}\ }\href {https://doi.org/10.1103/PhysRevLett.112.047004} {\bibfield  {journal} {\bibinfo  {journal} {Phys. Rev. Lett.}\ }\textbf {\bibinfo {volume} {112}},\ \bibinfo {pages} {047004} (\bibinfo {year} {2014})}\BibitemShut {NoStop}%
\bibitem [{\citenamefont {Ni}\ \emph {et~al.}(2020)\citenamefont {Ni}, \citenamefont {Karagoz}, \citenamefont {Gellman},\ and\ \citenamefont {P~De~Boer}}]{Ni2020boe}%
  \BibitemOpen
  \bibfield  {author} {\bibinfo {author} {\bibfnamefont {L.}~\bibnamefont {Ni}}, \bibinfo {author} {\bibfnamefont {B.}~\bibnamefont {Karagoz}}, \bibinfo {author} {\bibfnamefont {A.~J.}\ \bibnamefont {Gellman}},\ and\ \bibinfo {author} {\bibfnamefont {M.}~\bibnamefont {P~De~Boer}},\ }\bibfield  {title} {\enquote {\bibinfo {title} {Compression and decompression of structural tantalum films exposed to buffered hydrofluoric acid},}\ }\href@noop {} {\bibfield  {journal} {\bibinfo  {journal} {Journal of Micromechanics and Microengineering}\ }\textbf {\bibinfo {volume} {30}} (\bibinfo {year} {2020})}\BibitemShut {NoStop}%
\bibitem [{\citenamefont {M{\ifmmode\ddot{u}\else\"{u}\fi}ller}, \citenamefont {Cole},\ and\ \citenamefont {Lisenfeld}(2019)}]{Muller2019Oct}%
  \BibitemOpen
  \bibfield  {author} {\bibinfo {author} {\bibfnamefont {C.}~\bibnamefont {M{\ifmmode\ddot{u}\else\"{u}\fi}ller}}, \bibinfo {author} {\bibfnamefont {J.~H.}\ \bibnamefont {Cole}},\ and\ \bibinfo {author} {\bibfnamefont {J.}~\bibnamefont {Lisenfeld}},\ }\bibfield  {title} {\enquote {\bibinfo {title} {{Towards understanding two-level-systems in amorphous solids: insights from quantum circuits}},}\ }\href {https://doi.org/10.1088/1361-6633/ab3a7e} {\bibfield  {journal} {\bibinfo  {journal} {Rep. Prog. Phys.}\ }\textbf {\bibinfo {volume} {82}},\ \bibinfo {pages} {124501} (\bibinfo {year} {2019})}\BibitemShut {NoStop}%
\bibitem [{\citenamefont {Anthony-Petersen}\ \emph {et~al.}(2024)\citenamefont {Anthony-Petersen}, \citenamefont {Biekert}, \citenamefont {Bunker}, \citenamefont {Chang}, \citenamefont {Chang}, \citenamefont {Chaplinsky}, \citenamefont {Fascione}, \citenamefont {Fink}, \citenamefont {Garcia-Sciveres}, \citenamefont {Germond}, \citenamefont {Guo}, \citenamefont {Hertel}, \citenamefont {Hong}, \citenamefont {Kurinsky}, \citenamefont {Li}, \citenamefont {Lin}, \citenamefont {Lisovenko}, \citenamefont {Mahapatra}, \citenamefont {Mayer}, \citenamefont {{McKinsey}}, \citenamefont {Mehrotra}, \citenamefont {Mirabolfathi}, \citenamefont {Neblosky}, \citenamefont {Page}, \citenamefont {Patel}, \citenamefont {Penning}, \citenamefont {Pinckney}, \citenamefont {Platt}, \citenamefont {Pyle}, \citenamefont {Reed}, \citenamefont {Romani}, \citenamefont {Santana~Queiroz}, \citenamefont {Sadoulet}, \citenamefont {Serfass}, \citenamefont {Smith}, \citenamefont {Sorensen}, \citenamefont {Suerfu}, \citenamefont {Suzuki},
  \citenamefont {Underwood}, \citenamefont {Velan}, \citenamefont {Wang}, \citenamefont {Wang}, \citenamefont {Watkins}, \citenamefont {Williams}, \citenamefont {Yefremenko},\ and\ \citenamefont {Zhang}}]{Petersen2024stress-induced}%
  \BibitemOpen
  \bibfield  {author} {\bibinfo {author} {\bibfnamefont {R.}~\bibnamefont {Anthony-Petersen}}, \bibinfo {author} {\bibfnamefont {A.}~\bibnamefont {Biekert}}, \bibinfo {author} {\bibfnamefont {R.}~\bibnamefont {Bunker}}, \bibinfo {author} {\bibfnamefont {C.~L.}\ \bibnamefont {Chang}}, \bibinfo {author} {\bibfnamefont {Y.-Y.}\ \bibnamefont {Chang}}, \bibinfo {author} {\bibfnamefont {L.}~\bibnamefont {Chaplinsky}}, \bibinfo {author} {\bibfnamefont {E.}~\bibnamefont {Fascione}}, \bibinfo {author} {\bibfnamefont {C.~W.}\ \bibnamefont {Fink}}, \bibinfo {author} {\bibfnamefont {M.}~\bibnamefont {Garcia-Sciveres}}, \bibinfo {author} {\bibfnamefont {R.}~\bibnamefont {Germond}}, \bibinfo {author} {\bibfnamefont {W.}~\bibnamefont {Guo}}, \bibinfo {author} {\bibfnamefont {S.~A.}\ \bibnamefont {Hertel}}, \bibinfo {author} {\bibfnamefont {Z.}~\bibnamefont {Hong}}, \bibinfo {author} {\bibfnamefont {N.}~\bibnamefont {Kurinsky}}, \bibinfo {author} {\bibfnamefont {X.}~\bibnamefont {Li}}, \bibinfo {author} {\bibfnamefont
  {J.}~\bibnamefont {Lin}}, \bibinfo {author} {\bibfnamefont {M.}~\bibnamefont {Lisovenko}}, \bibinfo {author} {\bibfnamefont {R.}~\bibnamefont {Mahapatra}}, \bibinfo {author} {\bibfnamefont {A.}~\bibnamefont {Mayer}}, \bibinfo {author} {\bibfnamefont {D.~N.}\ \bibnamefont {{McKinsey}}}, \bibinfo {author} {\bibfnamefont {S.}~\bibnamefont {Mehrotra}}, \bibinfo {author} {\bibfnamefont {N.}~\bibnamefont {Mirabolfathi}}, \bibinfo {author} {\bibfnamefont {B.}~\bibnamefont {Neblosky}}, \bibinfo {author} {\bibfnamefont {W.~A.}\ \bibnamefont {Page}}, \bibinfo {author} {\bibfnamefont {P.~K.}\ \bibnamefont {Patel}}, \bibinfo {author} {\bibfnamefont {B.}~\bibnamefont {Penning}}, \bibinfo {author} {\bibfnamefont {H.~D.}\ \bibnamefont {Pinckney}}, \bibinfo {author} {\bibfnamefont {M.}~\bibnamefont {Platt}}, \bibinfo {author} {\bibfnamefont {M.}~\bibnamefont {Pyle}}, \bibinfo {author} {\bibfnamefont {M.}~\bibnamefont {Reed}}, \bibinfo {author} {\bibfnamefont {R.~K.}\ \bibnamefont {Romani}}, \bibinfo {author} {\bibfnamefont
  {H.}~\bibnamefont {Santana~Queiroz}}, \bibinfo {author} {\bibfnamefont {B.}~\bibnamefont {Sadoulet}}, \bibinfo {author} {\bibfnamefont {B.}~\bibnamefont {Serfass}}, \bibinfo {author} {\bibfnamefont {R.}~\bibnamefont {Smith}}, \bibinfo {author} {\bibfnamefont {P.}~\bibnamefont {Sorensen}}, \bibinfo {author} {\bibfnamefont {B.}~\bibnamefont {Suerfu}}, \bibinfo {author} {\bibfnamefont {A.}~\bibnamefont {Suzuki}}, \bibinfo {author} {\bibfnamefont {R.}~\bibnamefont {Underwood}}, \bibinfo {author} {\bibfnamefont {V.}~\bibnamefont {Velan}}, \bibinfo {author} {\bibfnamefont {G.}~\bibnamefont {Wang}}, \bibinfo {author} {\bibfnamefont {Y.}~\bibnamefont {Wang}}, \bibinfo {author} {\bibfnamefont {S.~L.}\ \bibnamefont {Watkins}}, \bibinfo {author} {\bibfnamefont {M.~R.}\ \bibnamefont {Williams}}, \bibinfo {author} {\bibfnamefont {V.}~\bibnamefont {Yefremenko}},\ and\ \bibinfo {author} {\bibfnamefont {J.}~\bibnamefont {Zhang}},\ }\bibfield  {title} {\enquote {\bibinfo {title} {A stress-induced source of phonon bursts and
  quasiparticle poisoning},}\ }\href {https://doi.org/10.1038/s41467-024-50173-8} {\bibfield  {journal} {\bibinfo  {journal} {Nature Communications}\ }\textbf {\bibinfo {volume} {15}},\ \bibinfo {pages} {6444} (\bibinfo {year} {2024})}\BibitemShut {NoStop}%
\bibitem [{\citenamefont {Turneaure}, \citenamefont {Halbritter},\ and\ \citenamefont {Schwettman}(1991)}]{Turneaure1991Oct}%
  \BibitemOpen
  \bibfield  {author} {\bibinfo {author} {\bibfnamefont {J.~P.}\ \bibnamefont {Turneaure}}, \bibinfo {author} {\bibfnamefont {J.}~\bibnamefont {Halbritter}},\ and\ \bibinfo {author} {\bibfnamefont {H.~A.}\ \bibnamefont {Schwettman}},\ }\bibfield  {title} {\enquote {\bibinfo {title} {{The surface impedance of superconductors and normal conductors: The Mattis-Bardeen theory}},}\ }\href {https://doi.org/10.1007/BF00618215} {\bibfield  {journal} {\bibinfo  {journal} {J. Supercond.}\ }\textbf {\bibinfo {volume} {4}},\ \bibinfo {pages} {341--355} (\bibinfo {year} {1991})}\BibitemShut {NoStop}%
\bibitem [{\citenamefont {Crowley}\ \emph {et~al.}(2023)\citenamefont {Crowley}, \citenamefont {McLellan}, \citenamefont {Dutta}, \citenamefont {Shumiya}, \citenamefont {Place}, \citenamefont {Le}, \citenamefont {Gang}, \citenamefont {Madhavan}, \citenamefont {Bland}, \citenamefont {Chang}, \citenamefont {Khedkar}, \citenamefont {Feng}, \citenamefont {Umbarkar}, \citenamefont {Gui}, \citenamefont {Rodgers}, \citenamefont {Jia}, \citenamefont {Feldman}, \citenamefont {Lyon}, \citenamefont {Liu}, \citenamefont {Cava}, \citenamefont {Houck},\ and\ \citenamefont {de~Leon}}]{Crowley2023disentangling_losses_Ta}%
  \BibitemOpen
  \bibfield  {author} {\bibinfo {author} {\bibfnamefont {K.~D.}\ \bibnamefont {Crowley}}, \bibinfo {author} {\bibfnamefont {R.~A.}\ \bibnamefont {McLellan}}, \bibinfo {author} {\bibfnamefont {A.}~\bibnamefont {Dutta}}, \bibinfo {author} {\bibfnamefont {N.}~\bibnamefont {Shumiya}}, \bibinfo {author} {\bibfnamefont {A.~P.~M.}\ \bibnamefont {Place}}, \bibinfo {author} {\bibfnamefont {X.~H.}\ \bibnamefont {Le}}, \bibinfo {author} {\bibfnamefont {Y.}~\bibnamefont {Gang}}, \bibinfo {author} {\bibfnamefont {T.}~\bibnamefont {Madhavan}}, \bibinfo {author} {\bibfnamefont {M.~P.}\ \bibnamefont {Bland}}, \bibinfo {author} {\bibfnamefont {R.}~\bibnamefont {Chang}}, \bibinfo {author} {\bibfnamefont {N.}~\bibnamefont {Khedkar}}, \bibinfo {author} {\bibfnamefont {Y.~C.}\ \bibnamefont {Feng}}, \bibinfo {author} {\bibfnamefont {E.~A.}\ \bibnamefont {Umbarkar}}, \bibinfo {author} {\bibfnamefont {X.}~\bibnamefont {Gui}}, \bibinfo {author} {\bibfnamefont {L.~V.~H.}\ \bibnamefont {Rodgers}}, \bibinfo {author} {\bibfnamefont
  {Y.}~\bibnamefont {Jia}}, \bibinfo {author} {\bibfnamefont {M.~M.}\ \bibnamefont {Feldman}}, \bibinfo {author} {\bibfnamefont {S.~A.}\ \bibnamefont {Lyon}}, \bibinfo {author} {\bibfnamefont {M.}~\bibnamefont {Liu}}, \bibinfo {author} {\bibfnamefont {R.~J.}\ \bibnamefont {Cava}}, \bibinfo {author} {\bibfnamefont {A.~A.}\ \bibnamefont {Houck}},\ and\ \bibinfo {author} {\bibfnamefont {N.~P.}\ \bibnamefont {de~Leon}},\ }\bibfield  {title} {\enquote {\bibinfo {title} {Disentangling losses in tantalum superconducting circuits},}\ }\href {https://doi.org/10.1103/PhysRevX.13.041005} {\bibfield  {journal} {\bibinfo  {journal} {Phys. Rev. X}\ }\textbf {\bibinfo {volume} {13}},\ \bibinfo {pages} {041005} (\bibinfo {year} {2023})}\BibitemShut {NoStop}%
\bibitem [{Gao Jiansong(2008)}]{Gao_Jiansong}%
  \BibitemOpen
  Gao Jiansong,\ \href {https://resolver.caltech.edu/CaltechETD:etd-06092008-235549} {\enquote {\bibinfo {title} {The physics of superconducting microwave resonators. dissertation (ph.d.), california institute of technology},}\ } (\bibinfo {year} {2008})\BibitemShut {NoStop}%
\bibitem [{\citenamefont {Mattis}\ and\ \citenamefont {Bardeen}(1958)}]{Mattis1985}%
  \BibitemOpen
  \bibfield  {author} {\bibinfo {author} {\bibfnamefont {D.~C.}\ \bibnamefont {Mattis}}\ and\ \bibinfo {author} {\bibfnamefont {J.}~\bibnamefont {Bardeen}},\ }\bibfield  {title} {\enquote {\bibinfo {title} {Theory of the anomalous skin effect in normal and superconducting metals},}\ }\href {https://doi.org/10.1103/PhysRev.111.412} {\bibfield  {journal} {\bibinfo  {journal} {Phys. Rev.}\ }\textbf {\bibinfo {volume} {111}},\ \bibinfo {pages} {412--417} (\bibinfo {year} {1958})}\BibitemShut {NoStop}%
\bibitem [{\citenamefont {Catelani}\ and\ \citenamefont {Pekola}(2022)}]{Catelani2022}%
  \BibitemOpen
  \bibfield  {author} {\bibinfo {author} {\bibfnamefont {G.}~\bibnamefont {Catelani}}\ and\ \bibinfo {author} {\bibfnamefont {J.~P.}\ \bibnamefont {Pekola}},\ }\bibfield  {title} {\enquote {\bibinfo {title} {Using materials for quasiparticle engineering},}\ }\href {https://doi.org/10.1088/2633-4356/ac4a75} {\bibfield  {journal} {\bibinfo  {journal} {Materials for Quantum Technology}\ }\textbf {\bibinfo {volume} {2}},\ \bibinfo {pages} {013001} (\bibinfo {year} {2022})}\BibitemShut {NoStop}%
\end{thebibliography}%


\begin{thebibliography}{9}%
\makeatletter
\providecommand \@ifxundefined [1]{%
 \@ifx{#1\undefined}
}%
\providecommand \@ifnum [1]{%
 \ifnum #1\expandafter \@firstoftwo
 \else \expandafter \@secondoftwo
 \fi
}%
\providecommand \@ifx [1]{%
 \ifx #1\expandafter \@firstoftwo
 \else \expandafter \@secondoftwo
 \fi
}%
\providecommand \natexlab [1]{#1}%
\providecommand \enquote  [1]{``#1''}%
\providecommand \bibnamefont  [1]{#1}%
\providecommand \bibfnamefont [1]{#1}%
\providecommand \citenamefont [1]{#1}%
\providecommand \href@noop [0]{\@secondoftwo}%
\providecommand \href [0]{\begingroup \@sanitize@url \@href}%
\providecommand \@href[1]{\@@startlink{#1}\@@href}%
\providecommand \@@href[1]{\endgroup#1\@@endlink}%
\providecommand \@sanitize@url [0]{\catcode `\\12\catcode `\$12\catcode `\&12\catcode `\#12\catcode `\^12\catcode `\_12\catcode `\%12\relax}%
\providecommand \@@startlink[1]{}%
\providecommand \@@endlink[0]{}%
\providecommand \url  [0]{\begingroup\@sanitize@url \@url }%
\providecommand \@url [1]{\endgroup\@href {#1}{\urlprefix }}%
\providecommand \urlprefix  [0]{URL }%
\providecommand \Eprint [0]{\href }%
\providecommand \doibase [0]{http://dx.doi.org/}%
\providecommand \selectlanguage [0]{\@gobble}%
\providecommand \bibinfo  [0]{\@secondoftwo}%
\providecommand \bibfield  [0]{\@secondoftwo}%
\providecommand \translation [1]{[#1]}%
\providecommand \BibitemOpen [0]{}%
\providecommand \bibitemStop [0]{}%
\providecommand \bibitemNoStop [0]{.\EOS\space}%
\providecommand \EOS [0]{\spacefactor3000\relax}%
\providecommand \BibitemShut  [1]{\csname bibitem#1\endcsname}%
\let\auto@bib@innerbib\@empty
\bibitem [{\citenamefont {Mueller}(1977)}]{MUELLER1977693}%
  \BibitemOpen
  \bibfield  {author} {\bibinfo {author} {\bibfnamefont {M.}~\bibnamefont {Mueller}},\ }\href {\doibase https://doi.org/10.1016/0036-9748(77)90141-7} {\bibfield  {journal} {\bibinfo  {journal} {Scripta Metallurgica}\ }\textbf {\bibinfo {volume} {11}},\ \bibinfo {pages} {693} (\bibinfo {year} {1977})}\BibitemShut {NoStop}%
\bibitem [{\citenamefont {Arakcheeva}\ \emph {et~al.}(2003)\citenamefont {Arakcheeva}, \citenamefont {Chapuis}, \citenamefont {Birkedal}, \citenamefont {Pattison},\ and\ \citenamefont {Grinevitch}}]{Arakcheeva:sn0032}%
  \BibitemOpen
  \bibfield  {author} {\bibinfo {author} {\bibfnamefont {A.}~\bibnamefont {Arakcheeva}}, \bibinfo {author} {\bibfnamefont {G.}~\bibnamefont {Chapuis}}, \bibinfo {author} {\bibfnamefont {H.}~\bibnamefont {Birkedal}}, \bibinfo {author} {\bibfnamefont {P.}~\bibnamefont {Pattison}}, \ and\ \bibinfo {author} {\bibfnamefont {V.}~\bibnamefont {Grinevitch}},\ }\href {\doibase 10.1107/S0108768103009005} {\bibfield  {journal} {\bibinfo  {journal} {Acta Crystallographica Section B}\ }\textbf {\bibinfo {volume} {59}},\ \bibinfo {pages} {324} (\bibinfo {year} {2003})}\BibitemShut {NoStop}%
\bibitem [{Gao Jiansong()}]{Gao_Jiansong}%
  \BibitemOpen
  Gao Jiansong,\ \href {https://resolver.caltech.edu/CaltechETD:etd-06092008-235549} {\enquote {\bibinfo {title} {The physics of superconducting microwave resonators. dissertation (ph.d.), california institute of technology},}\ } (\bibinfo {year} {2008})\BibitemShut {NoStop}%
\bibitem [{\citenamefont {Krupka}\ \emph {et~al.}(1999)\citenamefont {Krupka}, \citenamefont {Derzakowski}, \citenamefont {Tobar}, \citenamefont {Hartnett},\ and\ \citenamefont {Geyer}}]{Krupka1999May}%
  \BibitemOpen
  \bibfield  {author} {\bibinfo {author} {\bibfnamefont {J.}~\bibnamefont {Krupka}}, \bibinfo {author} {\bibfnamefont {K.}~\bibnamefont {Derzakowski}}, \bibinfo {author} {\bibfnamefont {M.}~\bibnamefont {Tobar}}, \bibinfo {author} {\bibfnamefont {J.}~\bibnamefont {Hartnett}}, \ and\ \bibinfo {author} {\bibfnamefont {R.~G.}\ \bibnamefont {Geyer}},\ }\href {\doibase 10.1088/0957-0233/10/5/308} {\bibfield  {journal} {\bibinfo  {journal} {Meas. Sci. Technol.}\ }\textbf {\bibinfo {volume} {10}},\ \bibinfo {pages} {387} (\bibinfo {year} {1999})}\BibitemShut {NoStop}%
\bibitem [{\citenamefont {Gr\"unhaupt}\ \emph {et~al.}(2018)\citenamefont {Gr\"unhaupt}, \citenamefont {Maleeva}, \citenamefont {Skacel}, \citenamefont {Calvo}, \citenamefont {Levy-Bertrand}, \citenamefont {Ustinov}, \citenamefont {Rotzinger}, \citenamefont {Monfardini}, \citenamefont {Catelani},\ and\ \citenamefont {Pop}}]{LukasG2018}%
  \BibitemOpen
  \bibfield  {author} {\bibinfo {author} {\bibfnamefont {L.}~\bibnamefont {Gr\"unhaupt}}, \bibinfo {author} {\bibfnamefont {N.}~\bibnamefont {Maleeva}}, \bibinfo {author} {\bibfnamefont {S.~T.}\ \bibnamefont {Skacel}}, \bibinfo {author} {\bibfnamefont {M.}~\bibnamefont {Calvo}}, \bibinfo {author} {\bibfnamefont {F.}~\bibnamefont {Levy-Bertrand}}, \bibinfo {author} {\bibfnamefont {A.~V.}\ \bibnamefont {Ustinov}}, \bibinfo {author} {\bibfnamefont {H.}~\bibnamefont {Rotzinger}}, \bibinfo {author} {\bibfnamefont {A.}~\bibnamefont {Monfardini}}, \bibinfo {author} {\bibfnamefont {G.}~\bibnamefont {Catelani}}, \ and\ \bibinfo {author} {\bibfnamefont {I.~M.}\ \bibnamefont {Pop}},\ }\href {\doibase 10.1103/PhysRevLett.121.117001} {\bibfield  {journal} {\bibinfo  {journal} {Phys. Rev. Lett.}\ }\textbf {\bibinfo {volume} {121}},\ \bibinfo {pages} {117001} (\bibinfo {year} {2018})}\BibitemShut {NoStop}%
\bibitem [{\citenamefont {Rieger}\ \emph {et~al.}(2023)\citenamefont {Rieger}, \citenamefont {G\"unzler}, \citenamefont {Spiecker}, \citenamefont {Nambisan}, \citenamefont {Wernsdorfer},\ and\ \citenamefont {Pop}}]{Rieger2023Fano}%
  \BibitemOpen
  \bibfield  {author} {\bibinfo {author} {\bibfnamefont {D.}~\bibnamefont {Rieger}}, \bibinfo {author} {\bibfnamefont {S.}~\bibnamefont {G\"unzler}}, \bibinfo {author} {\bibfnamefont {M.}~\bibnamefont {Spiecker}}, \bibinfo {author} {\bibfnamefont {A.}~\bibnamefont {Nambisan}}, \bibinfo {author} {\bibfnamefont {W.}~\bibnamefont {Wernsdorfer}}, \ and\ \bibinfo {author} {\bibfnamefont {I.}~\bibnamefont {Pop}},\ }\href {\doibase 10.1103/PhysRevApplied.20.014059} {\bibfield  {journal} {\bibinfo  {journal} {Phys. Rev. Appl.}\ }\textbf {\bibinfo {volume} {20}},\ \bibinfo {pages} {014059} (\bibinfo {year} {2023})}\BibitemShut {NoStop}%
\bibitem [{\citenamefont {Crowley}\ \emph {et~al.}(2023)\citenamefont {Crowley}, \citenamefont {McLellan}, \citenamefont {Dutta}, \citenamefont {Shumiya}, \citenamefont {Place}, \citenamefont {Le}, \citenamefont {Gang}, \citenamefont {Madhavan}, \citenamefont {Bland}, \citenamefont {Chang}, \citenamefont {Khedkar}, \citenamefont {Feng}, \citenamefont {Umbarkar}, \citenamefont {Gui}, \citenamefont {Rodgers}, \citenamefont {Jia}, \citenamefont {Feldman}, \citenamefont {Lyon}, \citenamefont {Liu}, \citenamefont {Cava}, \citenamefont {Houck},\ and\ \citenamefont {de~Leon}}]{Crowley2023disentangling_losses_Ta}%
  \BibitemOpen
  \bibfield  {author} {\bibinfo {author} {\bibfnamefont {K.~D.}\ \bibnamefont {Crowley}}, \bibinfo {author} {\bibfnamefont {R.~A.}\ \bibnamefont {McLellan}}, \bibinfo {author} {\bibfnamefont {A.}~\bibnamefont {Dutta}}, \bibinfo {author} {\bibfnamefont {N.}~\bibnamefont {Shumiya}}, \bibinfo {author} {\bibfnamefont {A.~P.~M.}\ \bibnamefont {Place}}, \bibinfo {author} {\bibfnamefont {X.~H.}\ \bibnamefont {Le}}, \bibinfo {author} {\bibfnamefont {Y.}~\bibnamefont {Gang}}, \bibinfo {author} {\bibfnamefont {T.}~\bibnamefont {Madhavan}}, \bibinfo {author} {\bibfnamefont {M.~P.}\ \bibnamefont {Bland}}, \bibinfo {author} {\bibfnamefont {R.}~\bibnamefont {Chang}}, \bibinfo {author} {\bibfnamefont {N.}~\bibnamefont {Khedkar}}, \bibinfo {author} {\bibfnamefont {Y.~C.}\ \bibnamefont {Feng}}, \bibinfo {author} {\bibfnamefont {E.~A.}\ \bibnamefont {Umbarkar}}, \bibinfo {author} {\bibfnamefont {X.}~\bibnamefont {Gui}}, \bibinfo {author} {\bibfnamefont {L.~V.~H.}\ \bibnamefont {Rodgers}}, \bibinfo {author} {\bibfnamefont
  {Y.}~\bibnamefont {Jia}}, \bibinfo {author} {\bibfnamefont {M.~M.}\ \bibnamefont {Feldman}}, \bibinfo {author} {\bibfnamefont {S.~A.}\ \bibnamefont {Lyon}}, \bibinfo {author} {\bibfnamefont {M.}~\bibnamefont {Liu}}, \bibinfo {author} {\bibfnamefont {R.~J.}\ \bibnamefont {Cava}}, \bibinfo {author} {\bibfnamefont {A.~A.}\ \bibnamefont {Houck}}, \ and\ \bibinfo {author} {\bibfnamefont {N.~P.}\ \bibnamefont {de~Leon}},\ }\href {\doibase 10.1103/PhysRevX.13.041005} {\bibfield  {journal} {\bibinfo  {journal} {Phys. Rev. X}\ }\textbf {\bibinfo {volume} {13}},\ \bibinfo {pages} {041005} (\bibinfo {year} {2023})}\BibitemShut {NoStop}%
\bibitem [{\citenamefont {Balashov}\ \emph {et~al.}(2018)\citenamefont {Balashov}, \citenamefont {Meyer},\ and\ \citenamefont {Wulfhekel}}]{Balashov2018}%
  \BibitemOpen
  \bibfield  {author} {\bibinfo {author} {\bibfnamefont {T.}~\bibnamefont {Balashov}}, \bibinfo {author} {\bibfnamefont {M.}~\bibnamefont {Meyer}}, \ and\ \bibinfo {author} {\bibfnamefont {W.}~\bibnamefont {Wulfhekel}},\ }\href {\doibase 10.1063/1.5043636} {\bibfield  {journal} {\bibinfo  {journal} {Rev. Sci. Instrum.}\ }\textbf {\bibinfo {volume} {89}},\ \bibinfo {pages} {113707} (\bibinfo {year} {2018})}\BibitemShut {NoStop}%
\bibitem [{\citenamefont {Murthy}\ \emph {et~al.}(2022)\citenamefont {Murthy}, \citenamefont {Lee}, \citenamefont {Kopas}, \citenamefont {Reagor}, \citenamefont {McFadden}, \citenamefont {Pappas}, \citenamefont {Checchin}, \citenamefont {Grassellino},\ and\ \citenamefont {Romanenko}}]{Murthy2022TOFSIMS}%
  \BibitemOpen
  \bibfield  {author} {\bibinfo {author} {\bibfnamefont {A.~A.}\ \bibnamefont {Murthy}}, \bibinfo {author} {\bibfnamefont {J.}~\bibnamefont {Lee}}, \bibinfo {author} {\bibfnamefont {C.}~\bibnamefont {Kopas}}, \bibinfo {author} {\bibfnamefont {M.~J.}\ \bibnamefont {Reagor}}, \bibinfo {author} {\bibfnamefont {A.~P.}\ \bibnamefont {McFadden}}, \bibinfo {author} {\bibfnamefont {D.~P.}\ \bibnamefont {Pappas}}, \bibinfo {author} {\bibfnamefont {M.}~\bibnamefont {Checchin}}, \bibinfo {author} {\bibfnamefont {A.}~\bibnamefont {Grassellino}}, \ and\ \bibinfo {author} {\bibfnamefont {A.}~\bibnamefont {Romanenko}},\ }\href {\doibase 10.1063/5.0079321} {\bibfield  {journal} {\bibinfo  {journal} {Applied Physics Letters}\ }\textbf {\bibinfo {volume} {120}},\ \bibinfo {pages} {044002} (\bibinfo {year} {2022})}\BibitemShut {NoStop}%
\end{thebibliography}%
\end{document}


\title{
    \textbf{Supplemental Material} \\ [0.7em]
    \Large \textbf{High quality superconducting tantalum resonators with beta phase defects}
}
\author[1]{\fontsize{10}{12}\selectfont Ritika~Dhundhwal}
\author[2]{Haoran~Duan}
\author[1]{Lucas~Brauch}
\author[1,3]{Soroush~Arabi}
\author[1]{Dirk~Fuchs}
\author[1]{Amir-Abbas~Haghighirad}
\author[4,5]{Alexander~Welle}
\author[6]{Florentine~Scharwaechter}
\author[6]{Sudip~Pal}
\author[6]{Marc~Scheffler}
\author[7]{Jos\'e~Palomo}
\author[7]{Zaki~Leghtas}
\author[8]{Anil~Murani}
\author[1,2]{Horst~Hahn}
\author[2]{Jasmin~Aghassi-Hagmann}
\author[2,5,9]{Christian~K\"ubel}
\author[1,3]{Wulf~Wulfhekel}
\author[1,3,6]{Ioan~M.~Pop}
\author[1]{Thomas~Reisinger}

\affil[1]{\itshape \mbox{IQMT, Karlsruhe Institute of Technology, 76131 Karlsruhe, Germany}}
\affil[2]{\itshape \mbox{INT, Karlsruhe Institute of Technology, 76131 Karlsruhe, Germany}}
\affil[3]{\itshape \mbox{PHI, Karlsruhe Institute of Technology, 76131 Karlsruhe, Germany}}
\affil[4]{\itshape \mbox{IFG, Karlsruhe Institute of Technology, 76131 Karlsruhe, Germany}}
\affil[5]{\itshape \mbox{KNMFi, Karlsruhe Institute of Technology, 76131 Karlsruhe, Germany}}
\affil[6]{\itshape \mbox{Physics Institute 1, University of Stuttgart, 70569 Stuttgart, Germany}}
\affil[7]{\itshape \mbox{Laboratoire de Physique de l’Ecole normale supérieure, 75005 Paris, France}}
\affil[8]{\itshape \mbox{Alice \& Bob, 53 Bd du Général Martial Valin, Paris 75015, France}}
\affil[9]{\itshape \mbox{Technische Universit\"at Darmstadt, 64289 Darmstadt, Germany}}
\date{}
\maketitle
\setcounter{figure}{0} 
\setcounter{table}{0}
\setcounter{equation}{0}
\renewcommand{\thefigure}{S\arabic{figure}} 
\renewcommand{\figurename}{Figure} 
\renewcommand{\thetable}{S\arabic{table}} 
\renewcommand{\tablename}{Table} 
\renewcommand{\theequation}{S\arabic{equation}} 

\section{Tantalum deposition}
Films nominally containing pure $\alpha$-phase were deposited at KIT, whereas film containing $\beta$-phase was grown at ENS Paris. We deposited approximately $\SI{200}{\nm}$ thick Ta films on 2" double-side-polished $c$-plane~(0001) sapphire using DC magnetron sputtering. The deposition parameters are summarized in~\cref{Table: sputtering_parameters}. For the single-phase films grown at KIT, we cleaned the substrates with Isopropanol (IPA) for $\SI{10}{\minute}$ in ultrasonic bath before loading them into the sputtering deposition system (model: Createc Fischer). The base pressure inside chamber was maintained at $10^{-9}$ mbar at room temperature. The substrates were baked overnight inside the high vacuum chamber at $\SI{300}{\degreeCelsius}$. We used a set of four halogen lamps to heat the substrate holder from the backside to $\SI{800}{\degreeCelsius}$, which corresponds to an actual temperature of around $\SI{530}{\degreeCelsius}$ on the sapphire surface. This was determined by attaching a thermocouple to the top surface of a loaded sapphire wafer in a separate calibration experiment. For the mixed-phase film grown at ENS Paris, we baked the substrate overnight inside the high vacuum chamber at $\SI{360}{\degreeCelsius}$ and cleaned it with Ar plasma for $\SI{150}{\s}$ inside the deposition chamber with following parameters: $\SI{20}{\sccm}$ Ar, $\SI{200}{\volt}$, $\SI{34}{\watt}$ and $\SI{0.018}{\m\bar}$.

\begin{table}[h]
\centering
\begin{tabular}{ m{4.5cm} p{2.1cm}  p{2.1cm}}
\toprule
    Parameter& single-phase\  &  mixed-phase \\ 
\toprule
    Substrate temperature$^*$ ($\si{\degreeCelsius}$)  &  800 &   650 \\ 
    Ar flow (sccm) &  44 &  13 \\  
    Pressure (mbar) & 0.003 & 0.009\\ 
    Power ($\si{\watt}$) &  200 & 500 \\ 
    Rate (nm/s) & 0.06  & 1 \\ 
    Cool-down time &  7 hours  &  3 hours \\ 
    Deposition site  & KIT & ENS Paris\\ 
\bottomrule
\end{tabular}
\caption{Magnetron sputtering parameters for Ta film deposition. 
$^*$ In case of KIT deposited films, substrate temperature was measured using a thermocouple not in contact with the substrate holder. To accurately determine the temperature of sapphire wafer for this temperature reading, we used a separate thermocouple attached to the sapphire wafer in a calibration experiment giving a calibrated temperature value of $\SI{530}{\degreeCelsius}$.}
\label{Table: sputtering_parameters} 
\end{table}

\section{XRD analysis of mixed-phase Ta}
We estimate the volume fraction of ${\beta}$-phase in mixed-phase film discussed in the text, S1 using the ratio of the integrated intensities of $\alpha$- and $\beta$-phase peaks of the XRD omega-scans, $I_{\alpha}$ and $I_{\beta}$, respectively, according to~\cref{eq:  volfraction}. We do not use the integrated peak intensities from the 2$\theta$-scans for this, as the thin film exhibits a fair amount of mosaicity. We look up the corresponding expected intensities, $I^{d}_{\alpha}$ and $I^{d}_{\beta}$, for powder diffraction in the Inorganic Crystal Structure Database (ICSD) selecting the following entries from it: ICSD code 652902~\cite{MUELLER1977693} for $\alpha$-phase and 54208~\cite{Arakcheeva:sn0032} for $\beta$-phase. Given the intensities listed there (see~\cref{table: beta_phase_vol_fraction}) and assuming the film is only composed of these two phases, the volume fraction is given by:
\begin{equation}
\frac{V_{\beta}}{V_{\alpha} + V_\beta} = \frac{I_{\beta} / I^d_{\beta}}{I_{\alpha}/I^{d}_{\alpha} + I_{\beta}/I^{d}_{\beta }}
\label{eq: volfraction}
\end{equation}
The volume fraction for the mixed-phase film determined using~\cref{eq: volfraction} is $9.9\%$. This is in good agreement with the amount of $\beta$-phase visible in the ACOM data (Fig.~1(b) of main text).

\begin{table}[h]
\centering
\begin{tabular}{p{1.5cm}p{1cm}p{3.5cm}p{3.7cm}}
\toprule
Peak   & $2\theta$   & Intensity (I, measured)    & Intensity ($I^d$, database) \\ 
\toprule
$\alpha$(110)   &  $38.32$   &  $145636$       & $100$   \\
$\beta$(002)    &  $33.68$   &  $4307$        & $26$    \\
\bottomrule
\end{tabular}
\caption{ICSD database intensities for the Ta phases and the intensities measured for the mixed-phase film.}
\label{table: beta_phase_vol_fraction}
\end{table}
We also determine the mosaicity of single- and mixed-phase films by performing rocking curve measurements (omega-scans) and use this parameter as an additional indication for the structural quality of the films. For mixed-phase film, we obtain film mosaicity of $\approx \SI{2}{\degree}$ and sub-degree mosaicity of $\SI{0.3}{\degree}$ for single-phase films.

\section{Resonator fabrication}
Samples S1 and S2 (as well as S5 and S6) were fabricated using e-beam lithography. Negative resist ma-N $2410$ (Micro Resist Technology GmbH) was spun on $7\times7 ~\si{\square\mm}$ Ta-coated sapphire wafers at $\SI{3000}{\rpm}$ for $\SI{60}{\s}$ and prebaked at $\SI{90}{\degreeCelsius}$ for $\SI{90}{\s}$. Resist coated wafers were exposed under e-beam with a dose of $\SI{350}{\micro\coulomb\per\cm\squared}$ using a Leo 1530, Raith ElphyPlus system. Exposed resist was developed in a TMAH based developer (Micro Resist Technology ma-D 525) for $\SI{3}{\minute}$ followed by dipping the wafers in a DI water stop-bath for $\SI{3}{\minute}$. To further harden the resist, the exposed wafers were postbaked at $\SI{100}{\degreeCelsius}$ for $\SI{15}{\minute}$ in an oven. The resonator pattern was transferred into the Ta film in a reactive ion etching(RIE) tool (SENTECH Instruments GmbH) using an SF$_6$ plasma with parameters: $\SI{25}{\sccm}$, $\SI{21}{\Pascal}$, $\SI{200}{\watt}$, $90$-$\SI{150}{\s}$. In order to improve the homogeneity of the etch we introduced a $\SI{30}{\s}$ oxygen descum prior to SF$_6$ etching with parameters: $\SI{15}{\sccm}$, $\SI{20}{\Pascal}$, $\SI{100}{\watt}$, $\SI{30}{\s}$. 

Samples S3 and S4 were fabricated using optical lithography. In this case, positive i-line resist S1805 (MICROPOSIT S1800 G2) was spun at $\SI{500}{\rpm}$ for $\SI{60}{\s}$ on the 2" Ta-coated wafers. Afterwards, we prebaked the wafers at $\SI{115}{\degreeCelsius}$ for $\SI{60}{\s}$. They were then exposed with $\SI{365}{\nm}$ ultra-violet mask aligner and developed in TMAH based developer Microposit MF $319$ for $\SI{30}{\s}$ followed by a $2$-$\SI{3}{\minute}$ DI water stop-bath. For transferring the pattern from resist to the Ta film, RIE was used with the same parameters as used for e-beam lithography process. At the end of the fabrication process, the wafers were diced into $7 \times 7$ mm$^{2}$ chips. 

\section{Temperature sweep fit results and discussion of kinetic inductance fraction}
For completeness, the parameters extracted from fitting the temperature dependence of frequency shift and internal quality factor ($Q_\text{i}$) are listed in the first four columns of~\cref{table:freq_fit_parameters}. The values for superconducting transition temperature~($T_\text{c}$) approximately agree when comparing the values extracted from thermal frequency shift and $Q_\text{i}$. The values are slightly lower for sample S1 compared to sample S3 and S4 and the bulk value of about $\SI{4.4}{\K}$. This is likely due to the well-proximitized inlcusions of the $\beta$-phase present in the film.

The resistivity of the film for sample S1 was also larger by almost an order of magnitude compared to the films used for samples S2-S4. The uncertainty of the fitted parameters is relatively large. We estimate it to be of the order of at least $10\%$, especially considering that the model equations are only rough approximations (see reference~\cite{Gao_Jiansong} for more details). Also the two parameters are relatively strongly correlated, although it can be stated that $T_\text{c}$ determines approximately the temperature where frequency or $Q_\text{i}$ deviations due to thermal quasi-particles start to dominate, and the kinetic inductance fraction ($\alpha_\text{k}$) is related to the slope. 

Note that the values for $A_\text{QP}^\text{Q(T)}$ and $\alpha_\text{k}^\text{f(T)}$ in the case of sample S2 deviate by two orders of magnitude, and one order of magnitude after buffered oxide etch (BOE) treatment. We interpret this as a small part of the resonator (approximately \SIrange{1}{10}{\percent} judging from the ratios in $\alpha_\text{k}$) having a depressed $T_\text{c}$ that dominates the generation of thermal quasi-particles at elevated temperature. 

In addition, the table lists measured ($f_0^{meas}$) and simulated ($f_0^{sim}$) frequencies of the resonators. The difference in the latter two frequency can be accounted for by assuming that it is due to the kinetic inductance neglected in the simulation model. Note that the $f_0^{meas}$ are very reproducible. The $f_0^{meas}$ of sample S1 is slightly lower due to the larger amount of kinetic inductance added by the high-resistivity $\beta$-phase.

For sample S2 the frequency shift does not fit with the high $\alpha_\text{k}$ extracted from temperature sweep data. This gives further support to the hypothesis that there is a small inhomogeneity with suppressed $T_\text{c}$ present in resonator S1. It can dominate the dependence on thermal quasi-particles due to the exponential dependence of their density on temperature, while the low temperature reactance is hardly affected.

Note that the $\alpha_\text{k}$ estimated in this way by~$\alpha_{\text{k}}^{\Delta\text{f}_0} = 1-(f_0^{meas}/f_0^{sim})^2$ is systematically larger than $\alpha_\text{k}^\text{f(T)}$. This is most likely both because of the simplified model used in the temperature sweep fits and remaining simplifications in the simulation. Possible sources of elevated capacitance are the high dielectric constant of Ta$_{2}$O$_{5}$ forming on the surface and sidewalls and the finite thickness of the metal sheets. In the simulation, we assumed anisotropic values for the relative permittivity of sapphire measured at $\SI{15}{\K}$, namely,  $\epsilon_{r,\parallel} = 11.34$ parallel to the c-axis and $\epsilon_{r,\perp} = 9.27$ orthogonal to it~\cite{Krupka1999May}. We also accounted for differences in fabrication most notably between e-beam and optical lithography, by matching the resonator pattern used in the simulation with optical microscopy images of the resonators.

\begin{table} 
\centering
\begin{tabular}{p{1.5cm}p{1.55cm}p{1.3cm}p{1.5cm}p{1.3cm}p{2cm}p{2cm}p{1.1cm}}
\toprule
Sample & $T_\text{c}^\text{Q(T)}$ (K) & $A_\text{QP}^\text{Q(T)}$ & $T_\text{c}^\text{f(T)}$ (K) & $\alpha_\text{k}^\text{f(T)}$ & $f_0^{meas}$ (GHz) & $f_0^{sim}$ (GHz) & $\alpha_{\text{k}}^{\Delta\text{f}_0}$  \\  
\toprule
S1       & 3.8  & 63.2 & 4.0  & 0.018   & 5.4315  & 5.62 & 0.066   \\ 
S1 (BOE) & 3.7  & 76.6 & 3.6 & 0.008   & 5.5079  & 5.62 & 0.040 \\     
S2       & 1  & 6065.6 & 1.1  & 0.0001  & 5.5124  & 5.62 & 0.038 \\ 
S2 (BOE) & 0.69 & 1121.1 & 0.88 & 0.0009  & 5.5091  & 5.62 & 0.039 \\ 
S3       & 4.6  & 23 & -  & -   & 5.3086  & 5.43 & 0.044 \\  
S3 (BOE) & 4.3  & 52.5 & 4.7  & 0.012   & 5.3087  & 5.43 & 0.044 \\
S4       & 4.4  & 50.1 & 4.8 & 0.011   & 5.3146  & 5.43 & 0.042 \\ 
S4 (BOE) & 4.4  & 58.6 & 4.7 & 0.009   & 5.3147  & 5.43 & 0.042 \\ 
\bottomrule
\end{tabular}
\caption{Summary of parameters resulting from fitting Eqs.~(1) and (2) of the main text to the temperature dependence of internal quality factor and frequency shift, as shown in Fig. 4(b,c) of main text, before and after buffered oxide etching, and comparison of the resulting kinetic inductance fraction ($\alpha_\text{k}^\text{f(T)}$) to that derived from frequency offset between measurement and simulation ($\alpha_{\text{k}}^{\Delta\text{f}_0}$). Here, "Sample" refers to the resonator samples discussed in the main text, and "BOE" indicates that the measurement was performed after buffered oxide etch treatment discussed in more detail in the next section. $T_\text{c}^\text{Q(T)}$ and 
$A_\text{QP}^\text{Q(T)}$ correspond to the values for superconducting critical temperature and a constant proportional to the inverse of the kinetic inductance fraction ($\alpha_\text{k}$), respectively, resulting from the fit to the thermal quasi-particle model in Eq.~(2). The values in columns $T_\text{c}^\text{f(T)}$ and 
$\alpha_\text{k}^\text{f(T)}$ refer to the fit results for superconducting critical temperature and ($\alpha_\text{k}$), respectively, using the model for the temperature dependent frequency shift in Eq.~(1). 
The column $f_0^{meas}$ lists the measured resonator frequencies at low power and base temperature. $f_0^{sim}$ refers to the resonator frequencies determined from simulation of the lumped element resonator in the waveguide using \emph{Ansys HFSS}, approximating the lumped element resonator to have zero thickness and neglecting any surface resistance and reactance. The kinetic inductance fraction estimated from these two frequencies is given by~$\alpha_{\text{k}}^{\Delta\text{f}_0} = 1-(f_0^{meas}/f_0^{sim})^2$.
}
\label{table:freq_fit_parameters}
\end{table}

\section{BOE post-processing} \label{BOE post-processing}
In this section we discuss the post-processing of the measured chips using BOE. After a description of the process we provide detailed resonator measurements and film characterization related to the BOE treatment.

For BOE treatment we dipped the sapphire chips with patterned Ta resonators in buffered hydrofluoric acid = BOE 7:1 (HF : NH$_4$F = 12.5 : 87.5\%) for $\SI{15}{\minute}$ at room temperature followed by 3 subsequent DI water baths and finally they were blown dry with nitrogen. Note that the chips containing resonators are mounted inside the 3D-waveguide sample holder by applying vacuum grease on the top and bottom edges of the chip. Therefore, for any post processing, chips were dismounted from the waveguide and were cleaned by dipping them in Xylene followed by acetone and IPA in ultrasonic bath to remove the vacuum grease. 

We do not observe any effect of the BOE treatment on the film structure in the XRD measurements on films that have not undergone lithographical patterning. The X-ray diffractograms for single- and mixed-phase films after BOE treatment are shown in~\cref{fig: xrd}.

\begin{figure*}
\centering
\includegraphics[width=0.75\linewidth]{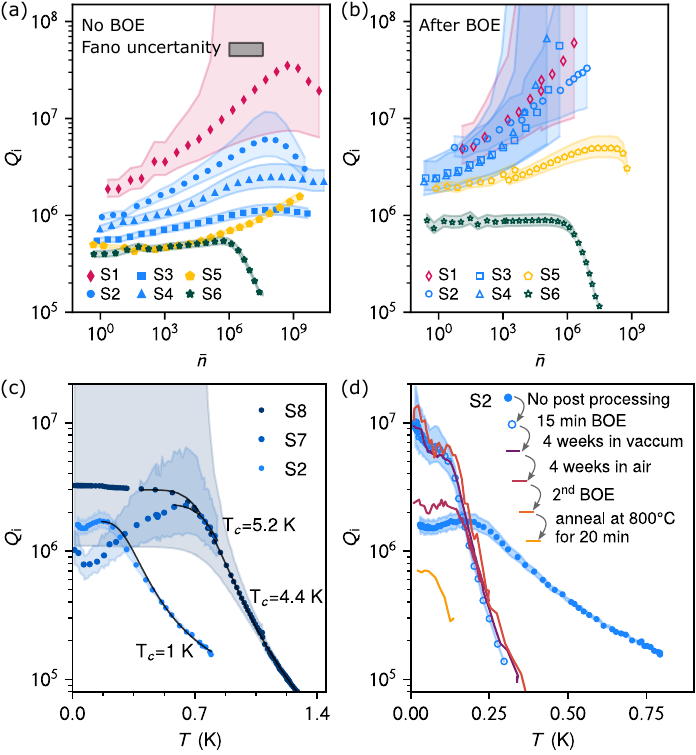}
\caption{Complete microwave characterization data of Ta resonators before and after buffered oxide etch (BOE) treatment, including test of stability and reproducibility of the improvement caused by BOE. (a) Internal quality factor ($Q_\text{i}$) as a function of average photon number $\bar{n}$ with Fano uncertainty shown as shaded intervals. Comparison of  $Q_\text{i}$ for mixed-phase (S1), [111] single-phase (S2-S4), and [110] single-phase Ta resonators (S5 and S6). S5 is deposited at KIT at a reduced substrate temperature of about $\SI{500}{\degreeCelsius}$ and S6 is a commercially (Star Cryoelectronics) deposited film. (b) $Q_\text{i}$ as a function of $\bar{n}$ for the same resonators after BOE treatment. (c) and (d) show $Q_\text{i}$ as a function of temperature for pure $\alpha$ Ta resonators. (c) shows three resonators, S2, S7 and S8 which are from different chips but same Ta wafer. S2 and S7 were measured at drive power of $\SI{-50}{\decibel}$  whereas S8 was measured at drive power of  $\SI{-20}{\decibel}$  resulting in a higher $Q_\text{i}$ than that of S2 and S7 in the low temperature regime. Superconducting transition temperature ($T_c$) is obtained   by fitting the decrease in $Q_\text{i}$ due to thermally induced quasi-particles in the high temperature regime to Eq.~(2) of the main text. (d) shows sample S2 before and after treating resonators in BOE repeatedly, different storage conditions between measurements, and after a final annealing step. For detailed description see text.
} 
\label{fig: Post-BOE-pwersweep}
\end{figure*}

\begin{figure}[h]
\centering
    \includegraphics{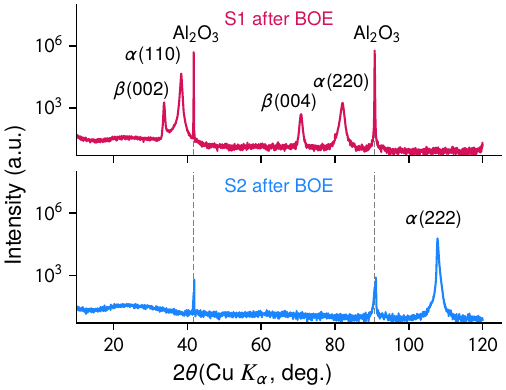}
    \caption{X-ray diffractograms of mixed- and single-phase Ta films after BOE treatment.}
    \label{fig: xrd}
\end{figure}

In~\crefadd{fig: Post-BOE-pwersweep}{(a,b)}, we characterized photon number dependence of $Q_\text{i}$ for Ta resonators before and after immersing them in BOE, additionally providing results for samples S5 and S6 and the complete power sweep data after the treatment. Samples S5 and S6 are Ta films with $\alpha$-phase growing with (110) orientation, in contrast to the single-phase films discussed in the main text that were composed of $\alpha$-phase growing with (111) orientation. S5 was deposited at KIT at a reduced substrate temperature of about $\SI{500}{\degreeCelsius}$ and S6 is a commercially deposited film from Star Cryoelectronics. \Crefadd{fig: Post-BOE-pwersweep}{(a)} shows $Q_\text{i}$ as a function of average photon number~\cite{LukasG2018} at the resonator ($\bar{n}$) with Fano uncertainty~\cite{Rieger2023Fano} shown in shaded area. As can be seen in~\crefadd{fig: Post-BOE-pwersweep}{(b)}, all resonators have $Q_\text{i}$ in the range $2-5 \times 10^6$ close to single photon regime except S6 after BOE treatment. The origin of this improvement remains an interesting puzzle with many possible candidates. We observe that more lossy resonators are improved by similar relative amounts which could either be interpreted as the same loss channel limiting these resonators or the BOE treatment affecting multiple loss channels. All resonators except S6 show dependence on $\bar{n}$. Improvement in $Q_\text{i}$ with $\bar{n}$ suggests that the saturable losses are still present after BOE treatment.

 In \crefadd{fig: Post-BOE-pwersweep}{(c)}, we measured $Q_\text{i}$ as a function of temperature for additional resonators, S7 and S8, from different chips but from the same single-phase Ta wafer as S2 and compare them with S2 (already shown in Fig. 4(c) of main text). These measurements were performed at fixed input drive power of $\SI{-50}{\decibel}$ for S2 and S7 and $\SI{-20}{\decibel}$ for S8. This explains the larger $Q_\text{i}$ for S8 in the low temperature regime. However, we do not expect a change in the $Q_\text{i}$ behavior in the high temperature regime due to input drive power and our comparison of thermally induced quasi-particle loss remains valid. We fit the $Q_\text{i}$ data with Eq.~(2) (see main text) based on the Mattis-Bardeen model with $T_c$ and $A_\text{QP}$ as free parameters. In case of S7 and S8, $T_c$ ($A_\text{QP}$) determined from the fit are $\SI{5.2}{\K}$ ($12.4$) and $\SI{4.4}{\K}$ ($50.4$), respectively.
Note that S7 was fabricated using same e-beam lithography based process used for S2 whereas S8 was fabricated using the optical lithography process.

To investigate the deviations in the temperature dependence of $Q_\text{i}$ for S2, we conduct a series of temperature sweep measurements on the same resonator in multiple cool-downs and treating it in different ways beforehand - see~\crefadd{fig: Post-BOE-pwersweep}{(d)}. 
After a first immersion in BOE, low temperature $Q_\text{i}$ improves but at elevated temperature the resonator becomes more susceptible to thermally induced quasi-particle loss, as already shown in the main text. Then, the same sample was kept in vacuum at room temperature within the cryostat for 4 weeks. After cooling it down without exposing it to air, the temperature dependence of $Q_\text{i}$ remains nearly the same (dark purple graph). After exposing it to ambient conditions for another 4 weeks, low temperature $Q_\text{i}$ is found to be almost same as before the BOE treatment (light purple graph). An improvement of nearly the same amount as after the first BOE treatment, could be achieved by immersing the same sample in BOE for second time (dark orange graph). After this second BOE treatment, the thermal quasi-particle induced loss at elevated temperature does not change. One of our hypothesis for this initial change of quality factor slope was a possible hydrogenation of the film during the BOE treatment. We test this by attempting to remove the hydrogen from the film in an annealing step in ultra-high vacuum ($\approx\SI{1e-7}{\milli\bar}$) at $\SI{800}{\degreeCelsius}$ for $\SI{20}{\minute}$. After this, we measured the sample again (dark yellow graph), resulting in a reduction of $Q_\text{i}$ both at low and intermediate temperature. A likely cause for this degradation is the diffusion of oxygen into the film,  due to the dissolution of the surface oxide and insufficient vacuum conditions, making the test for hydrogen inconclusive.

\begin{figure}
\centering
\includegraphics{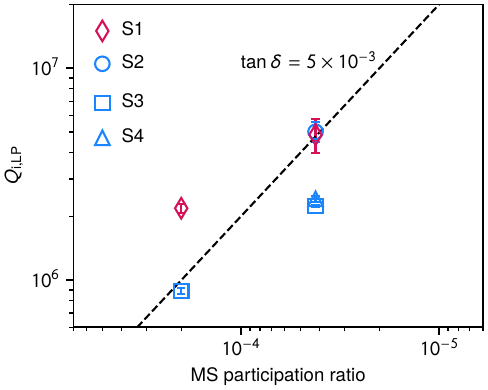}
\caption{Internal quality factor of BOE treated resonators at low power (LP) as a function of metal-substrate (MS) participation ratio. The dashed line shows the fit to the data and corresponds to a loss tangent~($\tan \delta$) of $5 \times 10^{-3}$}.
\label{fig: participation_ratio}
\end{figure}

In addition, we measured $Q_\text{i,LP}$, defined as $Q_\text{i}$ at low-power (LP), for a resonator design (capacitor gap of $\SI{10}{\um}$ instead of $\SI{50}{\um}$) with higher metal-substrate (MS) energy participation ratio ($p_\text{MS}$) in order to test if the reduction in $Q_\text{i,LP}$ is as expected for a dielectric-loss limited resonator. \Cref{fig: participation_ratio} shows the $Q_\text{i,LP}$ as a function of MS participation ratio for BOE treated resonators. We use the horseshoe (see main text) and interdigitated (see~\crefadd{fig: cryostat_wiring}{(c)}) style resonators with capacitor gaps of $\SI{50}{\um}$ and $\SI{10}{\um}$, respectively. By assuming an interface thickness of $\SI{3}{\nm}$ and a relative dielectric constant of $10$, we calculated the $p_\text{MS}$ using finite element simulation software \emph{Ansys HFSS}. The fit to the data results in a loss tangent ($\tan \delta$) of $5 \times 10^{-3}$ indicating that the resonators are limited by dielectric losses~\cite{Crowley2023disentangling_losses_Ta}. The deviations in $Q_\text{i,LP}$ from the fit could be a possible outcome of the different fabrication recipes used.

\section{ACOM and STEM}
The STEM characterization shown in Fig.~(1) and (3) of the main text was carried out on a Thermo Fischer Scientific Titan Themis 300 (S)TEM equipped with a Nanomegas ASTAR system and a Dectris Quadro camera. All experiments were performed at $\SI{300}{\kilo\volt}$ acceleration voltage in nanoprobe mode with $\SI{30}{\milli\radian}$ convergence for High angle annular dark field (HAADF) imaging and in microprobe precession mode at $\SI{1}{\milli\radian}$ convergence and $\SI{0.6}{\degree}$ precession angle for ACOM. Electron transparent cross sections of the thin films were prepared on a FEI Strata 400 S dual beam SEM/FIB. Protective layers of electron and ion beam deposited platinum were applied successively before trench cutting to preserve the original oxide layer structure.

\section{STM Measurement}
Scanning tunneling microscopy measurements presented in Fig.~2(b) of the main text were performed at a base temperature of $\SI{45}{\m\K}$ using a home-built dilution refrigerator setup~\cite{Balashov2018}. The superconducting gap and critical magnetic field of the films were characterized by $dI/dV$ measurements conducted in out-of-plane magnetic fields. These measurements were carried out with a bias voltage of $V_{b}=\SI{1.5}{\m\volt}$ and a current setpoint of $I_{p}= \SI{100}{\pico\ampere}$, utilizing a standard lock-in technique with an AC modulation voltage of $V_{\text{mod}}=\SI{20}{\micro\volt}$.

\section{DC Transport measurement}
We measured the resistance of Ta films as a function of temperature using the standard four-probe method using a Physical Property Measurement System (Quantum Design). The results are shown in~\cref{table: DC_transport}. For approx. $\SI{200}{\nm}$ film thickness, mixed- and single-phase films have $T_\text{c}$ of $\SI{4.2}{\K}$ and $\SI{4.3}{\K}$, respectively. While all films have $T_\text{c}$ close to the bulk value of $\SI{4.4}{\K}$, their residual resistance ratio (RRR) varies significantly. RRR is the ratio of resistance at room temperature and at the temperature just above $T_\text{c}$. It is a measure of the defects and impurities present in the film and is a gauge of the film quality. We observe lowest RRR of 4 in mixed-phase film, which is due to presence of the $\beta$-phase whereas single-phase films have highest RRR of 95. After BOE treatment, RRR either deteriorates as in case of S2 and S4 or remains same as in case of S1 and S3. The drastic decrease in RRR of S2 can be explained by the multiple BOE steps and annealing at $\SI{800}{\degreeCelsius}$ inside the high vacuum sputtering chamber for $\SI{20}{\minute}$ (see~\cref{BOE post-processing}). 
The lower $T_\text{c}$ of S2 after several steps of post-treatment was measured by directly contacting the meander inductor of the resonator. Therefore, it is unclear whether the change was due to different measurement conditions or post-treatment steps. 

\begin{table*}
\centering
\begin{tabular}{c|c|c|p{3cm}|c|c|p{4.5cm}}
\toprule
         Sample&  \multicolumn{3}{c|}{No BOE}&  \multicolumn{3}{c}{After BOE}\\\cline{2-7}  
         \addlinespace[0.4ex]
         & $T_\text{c}$ (K)&  RRR&  Sample state&  $T_\text{c}$ (K)&  RRR& Sample state\\ 
         \addlinespace[0.3ex]\toprule
         S1&  4.2&  4&  Diced chip before patterning &   4.1&   4& Patterned, contacted meander through wire bonds on capacitor plates of resonator \\ 
         S2&  4.3&  95&  Diced chip before patterning &  3.7&   2&  patterned, contacted meander through wire bonds on capacitor plates of resonator, BOE and annealed for $\SI{20}{\minute}$ \\
         S3&  4.3&  77&  Patterned, Hall bar &  4.3 &  77 & Patterned, Hall bar\\ 
         S4&  4.3&  93&  Patterned, Hall bar &  4.3 &  62 & Patterned, Hall bar \\
\bottomrule
\end{tabular}
\caption{DC transport measurements using four probe method: Superconducting transition temperature ($T_\text{c}$) and residual resistance ratio (RRR) of Ta films before and after buffer oxide etchant treatment.}
\label{table: DC_transport}
\end{table*}

\section{Microwave Measurement Setup}
\begin{figure}
\centering
\includegraphics[width=0.6\linewidth]{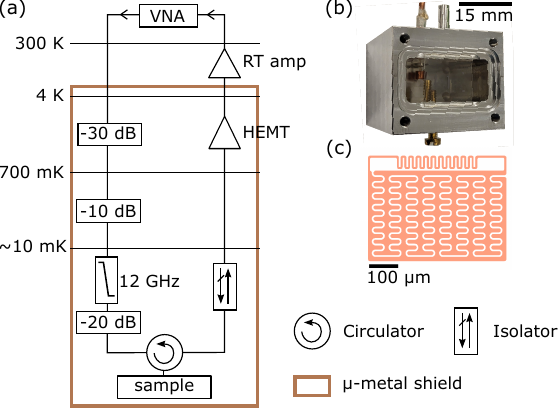}
\caption{(a) Microwave setup in dilution cryostat for resonator characterization. The input line to the waveguide is attenuated and filtered by a low-pass filter. Output line includes an isolator followed by an additional low-pass filter. The signal is then amplified using a commercial high electron mobility transistor amplifiers (HEMT) attached to the $\SI{4}{\K}$ stage and a room temperature (RT) amplifier. (b) Aluminum 3D-waveguide (without lid) used in the experiments with resonators (c) to vary the dielectric participation ratio. The gap between capacitor pads is $\SI{10}{\um}$.}
\label{fig: cryostat_wiring}
\end{figure}

The microwave measurements were carried out in a BlueFors dilution cryostat where the sample holder is attached to the mixing chamber plate of it. \Cref{fig: cryostat_wiring} shows the wiring diagram and various components of the cryostat. The outgoing microwave signal from a vector network analyzer (VNA) was attenuated by $60$-$\SI{70}{\dB}$ at various cryogenic stages before reaching the resonator through a circulator. The signal reflected from the resonator sample in the waveguide passes through an isolator at mixing chamber plate and was amplified using  a high-electron mobility transistor (HEMT) amplifier at $\SI{4}{\K}$ stage. The signal was further amplified at room temperature before it was recorded by the VNA.

\section{Time-of-Flight Secondary Ion Mass Spectrometry (ToF-SIMS)}
\begin{figure*}
    \centering
    \includegraphics[width=1\columnwidth]{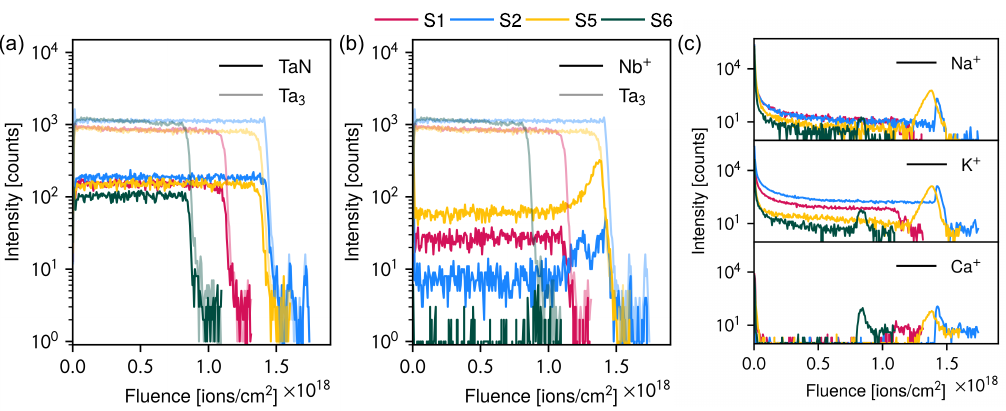}
    \caption{Time-of-Flight Secondary Ion Mass Spectrometry depth profile measurements on S1, S2, S4 and S5 Ta wafers (only diced, no lithography steps performed). Intensity of sputtered secondary ions is plotted against the fluence which is defined as number of impinged sputter ions per $\si{\cm}^2$ area. Zero fluence is associated with the film's surface and at the high-fluence end of the spectra is the metal-substrate interface. (a) shows the presence of TaN in all films. (b) shows a more distinct variation in the concentration of Nb$^+$ in the films. (c) shows the presence of contaminants like Na$^+$, K$^+$ and Ca$^+$ in the films.}
    \label{fig: TOF-SIMS}
\end{figure*}

We investigated samples S1, S2, S5 and S6 using ToF-SIMS, which is a surface analysis technique providing elemental and molecular information. Ta-coated sapphire wafers, originating from the same wafers as used for fabricating the resonators, were eroded by sputtering with Cesium or Oxygen ion sources under high vacuum. The ions removed from the film were mass separated and counted resulting in a mass spectrum as shown in~\cref{fig: TOF-SIMS}. The spectroscopy was performed on a ToF.SIMS5-100 (ION-TOF GmbH) at KMNFi, KIT. ToF-SIMS is a semi-quantitative and highly sensitive technique for characterizing impurity concentrations. Here we seek to correlate the impurities detected with resonator performance~\cite{Murthy2022TOFSIMS}. The impurities present in the Ta films in general depend on the purity level of sputtering target and sputtering gas, various physical and chemical treatments and handling during fabrication process and wafer dicing. \Cref{fig: TOF-SIMS} shows ToF-SIMS depth profile of the four films where secondary ion intensities are plotted against the sputter ion fluence [ions/cm\textsuperscript{2}] as an arbitrary measure of depth. In addition to elemental composition, the spectra also contain information about the relative thickness of the films. In our case, the order of films thickness in decreasing order is S2 and S5 (deposited at KIT) have identical thickness, S1 (ENS, Paris) and S6 (Star Cryoelectronics). Contaminants such as TaN, Nb$^+$, Na$^+$, K$^+$ and Ca$^+$ are present in all films. TaN is a common byproduct of magnetron sputtering and is present in all films as shown in~\crefadd{fig: TOF-SIMS}{(a)}. However, the total TaN content is similar in all samples when normalized with respect to Ta signal. Most importantly, magnetic materials such as iron and nickel are not found in the films. \Crefadd{fig: TOF-SIMS}{(b)} shows amount of Nb$^+$ found in the films. There is an increase in Nb$^+$ signal at the bottom of Ta film in S5, while the commercial sample S6 exhibits a nearly negligible amount of Nb. \Crefadd{fig: TOF-SIMS}{(c)} shows contaminants such as Na$^+$, K$^+$ and Ca$^+$ found both at the film surface and the bottom of the films and are a common result of water used in wafer polishing and dicing. Comparing this to the resonator quality factors we note that the commercially deposited Ta film S6 has the least amount of impurities overall but it does not perform best in microwave measurements. 

\hfill\break
\hfill\break
\hfill\break
\hfill\break
\hfill\break
\bibliographystyle{apsrev4-1} 
\bibliography{aipsamp_supplement}